\documentclass[fleqn,usenatbib]{mnras}

\usepackage{newtxtext,newtxmath}
\usepackage[flushleft]{threeparttable}
\usepackage[T1]{fontenc}

\DeclareRobustCommand{\VAN}[3]{#2}
\let\VANthebibliography\thebibliography
\def\thebibliography{\DeclareRobustCommand{\VAN}[3]{##3}\VANthebibliography}

\usepackage{graphicx}	
\usepackage{amsmath}	

\usepackage{color}
\definecolor{byzantine}{rgb}{0.74, 0.2, 0.64}

\usepackage{orcidlink}

\title[Mono-transits in TESS]{PLATO on the shoulders of TESS: analyzing mono-transit planet candidates in TESS data as a prior knowledge for PLATO observations}

\author[Magliano et al.]{Christian Magliano $^{1,2,3}$\thanks{E-mail: christian.magliano@unina.it}\orcidlink{0000-0001-6343-4744},
Giovanni Covone$^{1,2,3}$\orcidlink{0000-0002-2553-096X},
Valerio Nascimbeni$^{4,5}$\orcidlink{0000-0001-9770-1214},
Laura Inno$^{3,6}$\orcidlink{0000-0002-0271-2664},
Jose I. Vines $^{7}$\orcidlink{0000-0002-1896-2377},
\newauthor
Veselin Kostov,$^{8,9}$\orcidlink{0000-0001-9786-1031},
Stefano Fiscale$^{6,3}$\orcidlink{0000-0001-8371-8525},
Valentina Granata$^{5,4}$\orcidlink{0000-0002-1425-4541},
Marco Montalto$^{5,4}$\orcidlink{0000-0002-7618-8308},
Isabella Pagano$^{10}$\orcidlink{0000-0001-9573-4928},
\newauthor
Giampaolo Piotto$^{5,4}$\orcidlink{0000-0002-9937-6387},
Vito Saggese $^{1,2}$.
\\
$^{1}$Dipartimento di Fisica "Ettore Pancini", Università di Napoli Federico II, Napoli, Italy\\
$^{2}$INFN, Sezione di Napoli, Complesso Universitario di Monte S. Angelo,
Via Cintia Edificio 6, 80126 Napoli, Italy \\
$^{3}$INAF - Osservatorio Astronomico di Capodimonte,
via Moiariello 16, 80131 Napoli, Italy \\
$^{4}$INAF – Osservatorio Astronomico di Padova, vicolo dell’Osservatorio 5, 35122 Padova, Italy \\
$^{5}$Dipartimento di Fisica e Astronomia “Galileo Galilei”, Università degli Studi di Padova, vicolo dell’Osservatorio 3, 35122 Padova, Italy\\
$^{6}$Science and Technology Department, Parthenope University of Naples, Naples 80143 Italy \\
$^{7}$Instituto de Astronom\'ia, Universidad Cat\'olica del Norte, Angamos 0610, 1270709, Antofagasta, Chile\\
$^{8}$NASA Goddard Space Flight Center, 8800 Greenbelt Road, Greenbelt, MD 20771, USA\\
$^{9}$GSFC Sellers Exoplanet Environments Collaboration, USA\\
$^{10}$INAF - Osservatorio Astrofisico di Catania, Catania 95123 ,Italy\\
}

\date{Accepted XXX. Received YYY; in original form ZZZ}

\pubyear{2023}

\begin{document}
\label{firstpage}
\pagerange{\pageref{firstpage}--\pageref{lastpage}}
\maketitle

\newcommand\gio[1]{{\color{red}#1}}
\newcommand{\npcs}{38 }
\newcommand{\nfps}{10 }

\begin{abstract}
The Transiting Exoplanet Survey Satellite (TESS) and the upcoming PLATO mission (PLAnetary Transits and Oscillations of stars) represent two space-based missions with complementary objectives in the field of exoplanet science. While TESS aims at detecting and characterizing exoplanets around bright and nearby stars on a relative short-period orbit, PLATO will discover a wide range of exoplanets including rocky planets within the habitable zones of their stars.
We analyze mono-transit events in TESS data around stars that will or could be monitored by the PLATO mission, offering a unique opportunity to bridge the knowledge gap between the two missions and gain deeper insights into exoplanet demographics and system architectures. We found $48$ TESS mono-transit events around stars contained in the all-sky PLATO Input Catalog; of these, at least four will be imaged on the first long-pointing PLATO field, LOPS2. We uniformly vetted this sample to rule out possible false positive detections thus removing $10$ signals from the original sample. We developed an analytic method which allows us to estimate both the orbital period and inclination of a mono-transit planet candidate using only the shape of the transit. 
We derived the orbital period and inclination estimates for $30$ TESS mono-transit planet candidates. Finally, we investigated whether these candidates are amenable targets for a CHEOPS observing campaign. 
\end{abstract}




\section{Introduction}
\label{sec:intro}
The continuous advancements in exoplanet science have significantly expanded our understanding of the vast exoplanetary population, prompting the need for comprehensive surveys to unveil the intricate details of exoplanet demographics.
The TESS space-based mission \citep{Ricker2015}, launched by NASA in 2018, has been designed to conduct an all-sky survey, observing more than 200,000 bright stars across the sky during its two-year primary mission. TESS completed its primary mission in July 2020 and its primary extended mission in September 2022. Now it has just entered into the second extended mission which will last almost three years. Equipped with four wide-field cameras, it monitors the sky into 26 sectors and observes each sector for approximately 27 days. Using this observing strategy, TESS has enabled the detection and characterization of more than three hundred exoplanets, fostering our understanding of the diverse exoplanetary population. 
The drawback of TESS's nearly complete coverage of the sky is that it is more sensitive towards short period planets, which limits the opportunities to discover exoplanets orbiting the so-called habitable zone of Sun-like stars. 
When the orbital period $P$ is longer than the total observational baseline, the probability a planet transits only one time its star decreases as $P^{-5/3}$ \citep{YeeGaudi2008}.
Therefore, as most of the sky is continuously observed by TESS for less than 27 days the expected yield of mono-transit events detected by TESS is larger if compared with previous missions like Kepler and K2 \citep{Cooke2018,Villanueva2019}. According to the simulations by \cite{Cooke2018}, about $10\%$ of predicted TESS discoveries will be mono-transit candidates. 
Using a semianalytic technique, \cite{Villanueva2019} estimated TESS will find over one thousand mono-transit events, with 
about $90\%$ of the sample accessible to photometric and spectroscopic follow-up investigations from the ground \citep{Yao2019,Bayliss2020,Gill2020}. 
Taking into account the primary TESS mission extension, \cite{Cooke2019} updated the findings of their previous work.
Combining TESS Year 1 and Year 4 observations, they predicted a total of $140$ mono-transit events in the Southern Ecliptic Hemisphere. In addition, they claimed to recover $189$ duo-transit events during the time span between the TESS Year 1 and 4 campaigns. For these systems, it is possible to constrain their orbital period through the period aliasing technique \citep{Cooke2021}.

Mono-transits events are more challenging to confirm as bonafide planets than planets transiting two or more times within the observing time window. Since we are not able to fold multiple transits, their physical properties are poorly constrained and there is a higher probability of false positive detections from sporadic events. Nevertheless, these events are worthy of deeper scientific investigation because, if confirmed as true planet, they: i) would further enrich our understanding of exoplanetary systems \citep{Alibert2013}; ii) may be long period worlds orbiting the habitable zone of their host stars. 
Both of these crucial points would provide valuable insights into whether configurations as the Solar System are rare occurrences in the Galaxy or not (e.g., \citealt{Kipping2016}). This is important in particular considering the fact that  planets with optimal conditions for life might orbit K or G-type stars \citep[e.g.,][]{Covone2021}.
Different facilities spread all over the world could play a crucial role in recovering the TESS mono-transit events (e.g., \citealt{Gill2019}) thus enabling the confirmation of these events as genuine planets while also providing improved characterization of their orbital parameters. 

In this context, the ESA space-based mission PLATO \citep{Rauer2014}, slated for launch in 2026, will build upon the foundations laid by TESS and Kepler/K2 missions. 
PLATO aims at conducting a comprehensive survey of exoplanets, particularly targeting Earth-like planets around Sun-like stars. In fact, by leveraging on its precise photometry and long-duration observations, it will delve into the statistical properties of exoplanets, providing invaluable insights into their occurrence rates, orbital architectures and habitability potential.
Therefore, PLATO will provide a unique opportunity to fill the void left by TESS. The synergistic collaboration between TESS and PLATO would yield mutual benefits, as it allows for the optimal utilization of TESS data and enhances the observational capabilities of PLATO by providing prior knowledge of the mono-transit systems under scrutiny. In summary, mono-transit events in TESS data could represent a boost for the scientific return of the PLATO mission. 

In this paper, we focus our attention on mono-transit planet candidates detected by TESS around the stars that PLATO will monitor in the future. We propose an analytic approach to retrieve an estimate of the orbital period and inclination of these candidates, based on the shape of the transit. 
The outline of the paper is the following. In Sect. \ref{sec:methods} we describe the analytic technique used to retrieve an approximation of a mono-transit event's orbital period and inclination. Then, in Sect. \ref{sec:sample} we construct the sample of mono-transit TESS candidates orbiting those stars that could be also monitored by PLATO. We discuss the results obtained by applying this technique to our sample in Sect. \ref{sec:results}. Finally, we summarize our conclusions in Sect. \ref{sec:conclusions}.

\section{Methods}
\label{sec:methods}

The most challenging aspect when investigating mono-transit events is to retrieve the orbital period of the planet. The period is not only a key parameter to investigate the formation pathway of an exoplanet, but it is also crucial to schedule follow-up observations. A mono-transit event is by definition a single dip in the light curve of a given star, thus making the use of Lomb-Scargle periodogram (\citealt{Lomb1976,Scargle1982}) or periodic transit fitting algorithms (e.g. \citealt{Kovacs2002,Hippke2019}) completely useless. 
Indeed, the primary purpose of the aforementioned techniques is to identify periodic signals in time-series photometric data. 
Once a mono-transit event is detected, Bayesian methods usually offer a valuable tool to analyze them. By incorporating prior knowledge, such as stellar properties and observational constraints, Bayesian analysis can provide quite accurate and informative estimates of the planet's parameters (e.g., \citealt{Osborn2016,Incha2023}).

However, \cite{Seager2003} (S03, hereafter) showed that it is possible to relate the orbital parameters of a transiting planet with the stellar properties and the transit features. Specifically, when assuming a circular orbit and the absence of contamination in the measured star flux from other sources, there exists a relationship between the total transit duration $t_T$ of a planet and its orbital period $P$ given by:

\begin{equation}
        t_T=\dfrac{P}{\pi}\arcsin{\left(\dfrac{R_*}{a}\left[\dfrac{[1+\sqrt{\delta}]^2-[a\cos i/R_*]^2}{1-\cos ^2(i)}\right]^{1/2}\right)} \, ,
        \label{eq:t_T}
\end{equation}

where $a$ is the semi-major axis of the planet, $\delta$ is the transit depth when neglecting limb-darkening effects, $i$ the orbital inclination and $R_*$ the stellar radius. Using Kepler's third law and assuming that the mass of the planet is small (i.e., $M_p\ll M_*$, where $M_*$ is the stellar mass), equation \eqref{eq:t_T} can be written as
\begin{equation}
        t_T-\dfrac{P}{\pi}\arcsin{\left[\left(\dfrac{\Lambda^{-2/3}_{*}P^{-4/3}(1+\sqrt{\delta})^2-\cos^2(i)}{1-\cos^2(i)}\right)^{1/2}\right]}=0 \, ,
        \label{eq:t_T_2}
\end{equation}
where we defined $\Lambda_{*}\equiv G\rho_{*}/3\pi$ with $\rho_*=3M_*/4\pi R_*^3$ the mean stellar density\footnote{Note we used a slightly different definition of $\rho_*$ with respect to S03.} and $G$ the gravitational constant.
Under the approximation $\pi t_T/P\ll 1$ and $\cos i\ll 1$ (which are reasonable for likely long-period transiting planets) equation \eqref{eq:t_T_2} can be cast in the following form:

\begin{equation}
    \cos^2(i) \, P^2-\Lambda^{-2/3}_{*}(1+\sqrt{\delta})^2 \, P^{2/3}+\pi^2 \, t_{T}^2=0 \, .
    \label{eq:t_T_3}
\end{equation}

Equations \eqref{eq:t_T_2}, or its approximated form \eqref{eq:t_T_3}, can be used to estimate the orbital period of mono-transit events (e.g., \citealt{YeeGaudi2008, Wang2015,Uehara2016}). For a fixed value of the orbital inclination $i$, if the mean stellar density of the host star $\rho_*$ is calculated from an independent observation, one can solve equations \eqref{eq:t_T_2} and \eqref{eq:t_T_3} in terms of the orbital period $P$. 
Equation \eqref{eq:t_T_2} is a transcendental equation for which only numeric solving approaches are possible. 
Equation \eqref{eq:t_T_3}, instead, is a quadratic and fractional power equation, so that an analytic resolution can be derived (see Sect. \ref{sec:polynomial_eq}). 
The uncertainty in the solutions to equations \eqref{eq:t_T_2} and \eqref{eq:t_T_3} will be the result of the combination of the uncertainty in the stellar density and the orbital parameters (see equations A18-A24 in Appendix \ref{appendix: formal_solutions} for details on the error propagation).
In general, for $i\approx 90^\circ$ the orbital period of a mono-transit event in a star's light curve can be approximated either through the numerical solution of equation \eqref{eq:t_T_2} or through the analytical solution of equation \eqref{eq:t_T_3}.
For $i$ exactly $90^\circ$ , i.e. in the edge-on (EO, $i=90^\circ$) condition, the quadratic term in the equation \eqref{eq:t_T_3} vanishes and the orbital period $P_{\text{EO}}$ is given by:

\begin{equation}
    P_{\text{EO}}=\left(\dfrac{\pi \, t_T}{1+\sqrt{\delta}}\right)^3\Lambda_* \, .
    \label{eq:p_eo}
\end{equation}

Equation \eqref{eq:p_eo} shows that the accuracy to which we can determine the orbital period depends on the accuracy on which we measure the mean stellar density \citep{Wang2015}. 
By combining Gaia parallaxes with isochrone fitting, \cite{Sandford2019} estimated the stellar densities for a selection of K2 stars hosting planets. Using the third Kepler's law to infer the orbital period, their analysis resulted in a fractional uncertainty of $15_{-6}^{+30}$ per cent for the orbital period, assuming a circular orbit.

However, the PLATO's asteroseismology programme aims at reducing the uncertainty on stellar mean density determinations to about $10\%$ for G0V stars with a visual magnitude of 10 \citep{Goupil2017}.
In light of this, the estimate of the orbital period could be updated using cutting-edge estimates of the stellar parameters.
Finally, as we will show in Sect. \ref{sec:polynomial_eq}, even a slight change on the orbital inclination can lead to very different values for the period. Consequently, the edge-on assumption can be considered a first-order approximation of the orbital period of a mono-transit planet candidate.

\subsection{Polynomial equation}
\label{sec:polynomial_eq}
For now, let us assume that the inclination of the orbit is known.
In Sect. \ref{sec:uncertainties} we will discuss the realistic scenario where the orbital inclination is an additional unknown in the equations \eqref{eq:t_T_2} and \eqref{eq:t_T_3}.
Let $\alpha$, $\beta$ and $\gamma$ be defined as
\begin{equation}
    \alpha\equiv \cos^2(i),\\
    \beta\equiv \Lambda^{-2/3}_{*}(1+\sqrt{\delta})^2, \\
    \gamma\equiv \pi^2 t_{T}^2 \, .
\end{equation}
Then equation \eqref{eq:t_T_3} can be written as
\begin{equation}
    f(P)\equiv \, \alpha \, P^2-\beta \, P^{2/3} + \gamma=0 \, .
    \label{eq:eq_param}
\end{equation}
For a transiting planet, equation \eqref{eq:eq_param} has two solutions $P_1$ and $P_2$ (degeneracy issue, see Sect. \ref{sec:degeneracy}), given by:

\begin{equation}    P_1=2^{3/2}\left(\dfrac{\beta}{3\alpha}\right)^{3/4}\cos^{3/2}\Bigg[\dfrac{5}{3}\pi-\dfrac{1}{3}\arctan\left(\sqrt{\dfrac{4\beta^3}{27\alpha\gamma^2}-1}\right)\Bigg],
    \label{eq:P1}
\end{equation}
\begin{equation}
    P_2=2^{3/2}\left(\dfrac{\beta}{3\alpha}\right)^{3/4}\cos^{3/2}\Bigg[\dfrac{\pi}{3}-\dfrac{1}{3}\arctan\left(\sqrt{\dfrac{4\beta^3}{27\alpha\gamma^2}-1}\right)\Bigg] \, .
    \label{eq:P2}
\end{equation}

The comprehensive calculations are available in the Appendix \ref{appendix: formal_solutions}. Moreover, equations \eqref{eq:P1} and \eqref{eq:P2} have physical meaning only if the argument of the square root is non negative, that is 

\begin{equation}
   \dfrac{2\sqrt{3}}{3\pi}\dfrac{(1+\sqrt{\delta})^3}{G \, t_T^2 \, \rho_* \, \cos i}\geq 1 \, .
    \label{eq:condition}
\end{equation}

As demonstrated in Appendix \ref{appendix: formal_solutions}, when $(4\beta^3/27\alpha\gamma^2)<1 $, equation \eqref{eq:eq_param} does not yield any physically meaningful solutions, while in the case of $(4\beta^3/27\alpha\gamma^2)=1$, only one solution is obtained.

\begin{figure*}
    \centering
    \includegraphics[width=\textwidth]{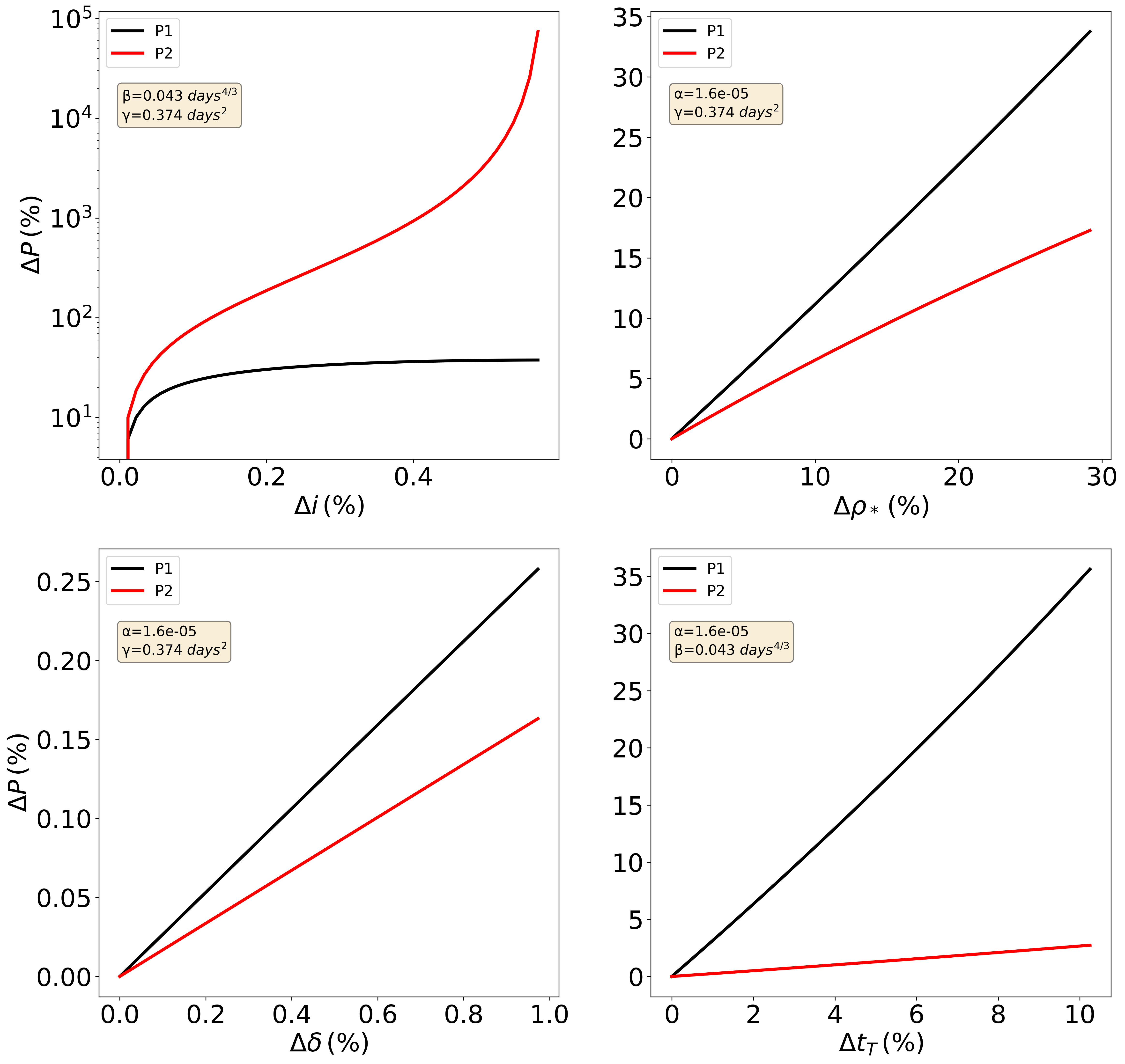}
    \caption{Variation of the solutions to equation \eqref{eq:eq_param}. On \textit{top left} we kept $\beta$ and $\gamma$ constant while varying $\alpha$ through a slight adjustment in the orbital inclination $i$ by a few percentage points. In the \textit{top right} and \textit{bottom left} plots, we fixed $\alpha$ and $\gamma$ steady while modifying $\beta$ by respectively altering the mean stellar density $\rho_*$ (top right) and adjusting the transit depth $\delta$ (bottom left). On \textit{bottom right} we fixed $\alpha$ and $\beta$ while varying $\gamma$ by changing the total duration of the transit $t_T$.}
    \label{fig:simpl_eq_behav}
\end{figure*}

In order to understand how changing the three coefficients affects the solutions to equation \eqref{eq:eq_param}, in Fig. \ref{fig:simpl_eq_behav} we show the relative variation of the two solutions $P_1$ and $P_2$ against the variation of one of the fundamental parameters.
We examined four specific scenarios, in each one we made a small adjustment to one of the parameters $i$, $\rho_*$, $\delta$, or $t_T$, 
while keeping the other three constant. To make our simulations as realistic as possible, for each parameter, we have considered the maximum variation to be the median relative percentage error observed in the TESS photometric data publicly available on the ExoFOP archive\footnote{https://exofop.ipac.caltech.edu/tess/}. In particular, we defined the TESS relative percentage error on each of the aforementioned parameters as the median of the distribution in order to damp the effects of the outliers.

As we can see in Fig. \ref{fig:simpl_eq_behav}, the orbital inclination (or $\alpha \equiv \cos^2 i$) is the primary physical parameter to which the function $f(P)$  exhibits the highest sensitivity. In fact, even if the inclination undergoes a small variation ($\Delta i<1\%$), $P_1$ and $P_2$ can span a broad spectrum of values. In this specific case, while the range of $P_1$ extends over a few dozen of percentages due to a $\Delta i$ up to $0.5\%$,  $P_2$ covers five orders of magnitudes with the same variation. The second parameter to which $f(P)$ is most sensitive is the duration of the transit $t_T$ (or $\gamma$) which causes a rigid translation upwards (if $t_T$ increases) or downwards (if $t_T$ decreases) of $f(P)$. TESS's typical $10\%$ variation in $t_T$ leads to an approximate $30\%$ change in $P_1$, while $P_2$ experiences a fluctuation of a few percentage points. 
Furthermore, we observed that a $30\%$ change in stellar density $\rho_{*}$ corresponds to a $35\%$ variation in $P_1$ and a $15\%$ variation in $P_2$. It is noteworthy that the $10\%$ accuracy on $\rho_*$ that will be achieved by PLATO's asteroseismology analysis would lead to a reduction of $10\%$ and $5\%$ in the variation of $P_1$ and $P_2$, respectively.
Finally, we remark that the solution $P_1$ and $P_2$ exhibit the lowest sensitivity to the variations in transit depth $\delta$. 
A typical $1\%$ variation in $\delta$ results in a minor change of a few tenths of a percentage in the period.
Fig. \ref{fig:simpl_eq_behav} clearly shows that $P_2$ is more robust than $P_1$ against variations of $\rho_*$, $\delta$ and $t_T$, however it is highly affected by tiny changes in the orbital inclination which dominates the uncertainty on $P_2$.

\begin{figure}
    \centering
    \includegraphics[width=0.47\textwidth]{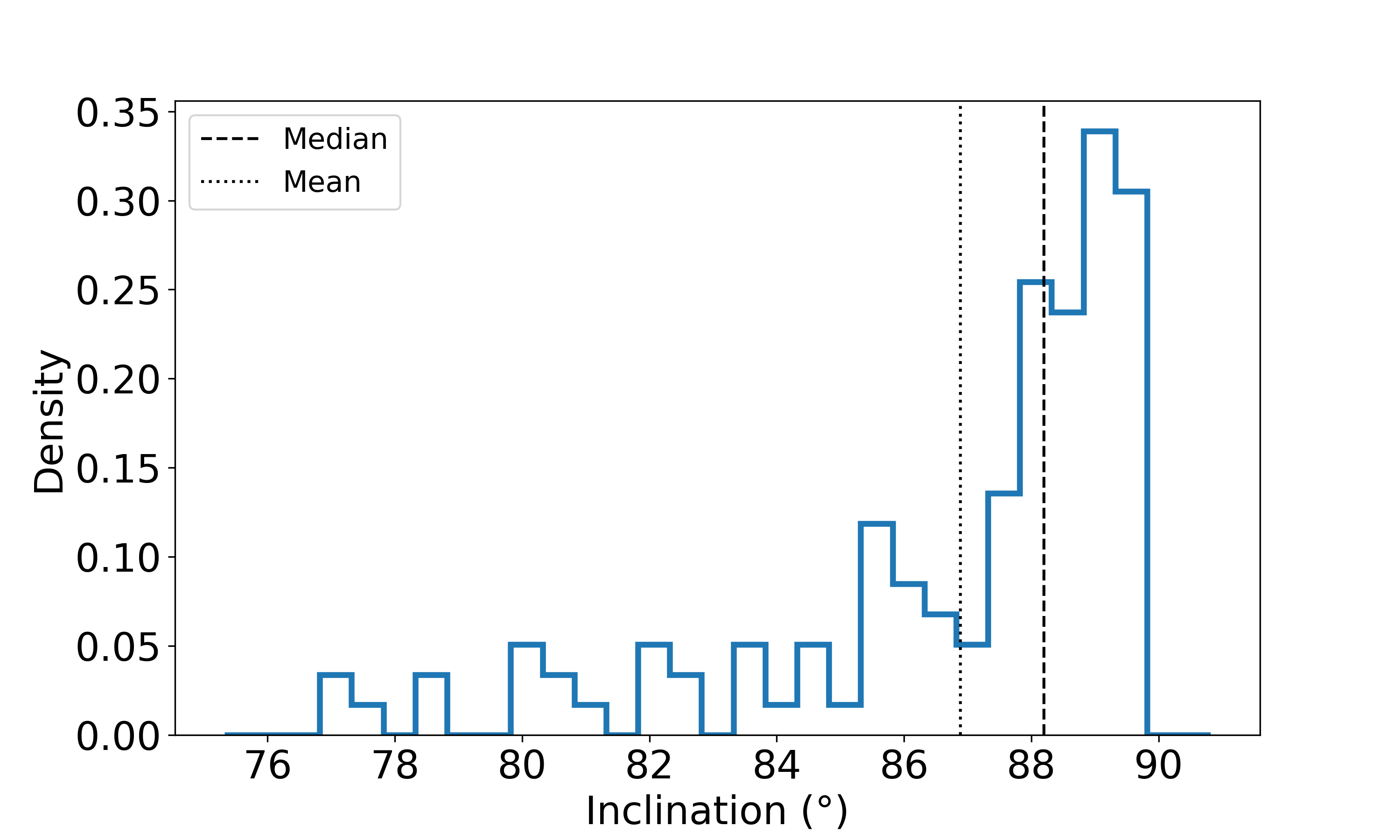}
    \caption{Distribution of TESS confirmed exoplanets according to their orbital inclination $i$. The dotted and dashed vertical lines represent respectively the mean ($87^\circ$) and the median ($88.2^\circ$) of the distribution.}
    \label{fig:tess_incl_hist}
\end{figure}

This section focuses on solving the polynomial equation \eqref{eq:t_T_3}, but the findings are also qualitatively valid for equation \eqref{eq:t_T_2}'s solutions since the condition $\cos i\ll 1$ is basically satisfied for all transiting planets. In Fig. \ref{fig:tess_incl_hist} we show the distribution of TESS confirmed exoplanets according to their orbital period. As we can see from the histogram, about $ 70\%$ of the total sample has inclinations within $87^\circ\leq i\leq 90.5^\circ$. 

Throughout our discussion we ignored that the limb-darkening affects the shape of the transit:  i) it affects the transit depth $\delta$ as a function of the orbital inclination $i$, ii) it reduces the elapsed time from the second to the third contact ($t_F$, hereafter) by rounding the flat bottom part of the transit and iii) it darkens the distinction between the ingress/egress phases and the flat bottom region. 
Finally, we are assuming that the light curve is not contaminated by other sources. However, since the big TESS pixel scale, this assumption may not be always met (more on this in Sect.\ref{sec:vetting}). Thus the findings of this method also rely on the angular resolution of the telescope used for the observation.

\subsection{The orbital inclination issue}
\label{sec:uncertainties}
Despite the majority of transiting exoplanets having almost $i=90^\circ$, a very small discrepancy ($\Delta i \approx 0.5\%$) in the orbital inclination can result in a drastic change in the orbital period solutions to equation \eqref{eq:eq_param} (as shown in Sect. \ref{sec:polynomial_eq}). 
In general, when the orbital inclination is close to $90^\circ$ the exoplanet's transit across the stellar disk is longer since it covers a larger distance. This leads to a longer transit duration. On the contrary, when the inclination deviates from $90^\circ$, the exoplanet's transit is shorter as it covers a smaller distance, resulting in a shorter transit duration. Moreover, as mentioned above, the orbital inclination would also affect the transit duration when accounting for the limb-darkening effect. 
Therefore, we assume here that the errors on the orbital parameters $\delta$ and $t_T$  depend only on TESS photometric precision and also the lack of information on the orbital inclination of the system under examination assuming the limb-darkening effect is negligible. 
The orbital inclination of a mono-transit planet candidate is not known because a mono-transit event only provides information about the total duration of the transit $t_T$ and the transit depth $\delta$. Therefore, in order to consider the orbital inclination as an additional unknown in equation \eqref{eq:t_T_3}, we will solve it for a grid of possible inclination values such that the equation \eqref{eq:condition} is satisfied, that is 
\begin{equation}
    i\leq \arccos\left(\dfrac{2\sqrt{3}}{3\pi}\dfrac{(1+\sqrt{\delta})^3}{G \, t_T^2 \, \rho_*}\right).
    \label{eq:grid}
\end{equation}
Thus we obtain two solutions, $P_1(i)$ and $P_2(i)$, as a function of the orbital inclination that describes all the possible combinations $(P, i)$ of the mono-transit event within the error bars.

\subsection{Degeneracy of the solutions}
\label{sec:degeneracy}

As observed in Sect. \ref{sec:polynomial_eq}, equations \eqref{eq:t_T_2} and \eqref{eq:t_T_3} may admit two solutions leading to a degeneracy. A possible way to break this degeneracy comes from the relationship that exists between the total transit duration $t_T$ and the elapsed time from the second to the third contact $t_F$, in the non-limb-darkening assumption (equations (9) and (7) of S03) and considering $M_p\ll M_*$, that is

\begin{equation}
    \dfrac{M_*}{R_*^3}=\left(\dfrac{4\pi^2}{P^2G}\right)\Biggl\{\dfrac{(1+\sqrt{\delta})^2-b^2(1-\sin^2(t_T\pi/P)}{\sin^2(t_T\pi/P)}\Biggr\}^{3/2} \, ,
    \label{eq:complex_eq}
\end{equation}
where
\begin{equation}
    b=\sqrt{\dfrac{(1-\sqrt{\delta})^2-\left[\sin^2(t_F\pi/P)/\sin^2(t_T\pi/P)\right](1+\sqrt{\delta})^2}{1-\left[\sin^2(t_F\pi/P)/\sin^2(t_T\pi/P)\right]}} \, .
    \label{eq:complex_eq_2}
\end{equation}
If we can measure $t_F$ from the transit shape, then we are able to numerically resolve equation \eqref{eq:complex_eq} in terms of the period $P$ for each value of the orbital inclination in the given grid. Thus from a theoretical point of view, equations \eqref{eq:t_T_3} (or eq.\eqref{eq:t_T_2}) and \eqref{eq:complex_eq} form a system of equations that allows us to break the degeneracy of the solutions.

S03 also showed that equation \eqref{eq:complex_eq} can take a simpler form under the assumption $R_*\ll a$, or equivalently $\pi \, t_T/P\ll 1$, that is always satisfied for long-period planets. In this approximation the orbital period $P$ is given by
\begin{equation}
    P=\dfrac{G\pi^2}{24}\rho_*\dfrac{(t_T^2-t_F^2)^{3/2}}{\delta^{3/4}}.
    \label{eq:simple_eq}
\end{equation}
Hereafter, we will refer to $P_{t_F}$ when the orbital period of the mono-transit planet candidate is calculated by means of the equation \eqref{eq:simple_eq}. While $P_1(i)$ and $P_2(i)$ solutions depend on the orbital inclination, $P_{t_F}$ represents an independent estimate which theoretically breaks the $P_1-P_2$ degeneracy. However, if the uncertainty $\Delta P_{t_F}$ over $P_{t_F}$ is large enough, its estimate could be compatible with both $P_1(i)$ and $P_2(i)$ branches. In Sect. \ref{sec:test}, we will demonstrate how equation \eqref{eq:simple_eq} helps in constraining the orbital period within a range of values and also allows to establish limits on the physically possible values of the orbital inclination.

We note that, even if we did not measure $t_F$, by putting $t_F=0$, equation \eqref{eq:simple_eq} gives the upper limit for the orbital period $P_{\text{max}}$, that is

\begin{equation}
    P_{\text{max}}=\dfrac{G\pi^2}{24}\rho_*\dfrac{t_T^{3}}{\delta^{3/4}} \, .
    \label{upper_period}
\end{equation}
Despite $t_F=0$ being a more theoretical limit than a realistic scenario, grazing transiting planets are such that $t_F/t_T\ll 1$, so $P_{\text{max}}$ represents a conservative estimate of the true maximum orbital period of the planet. Using equation \eqref{upper_period}, \cite{Cacciapuoti2002} estimated the upper period of the candidate planet TOI-411 d to be $46 \pm 4$ days. Based on the combined TESS and CHEOPS observations, \cite{Garai2023} determined the true period to be $47.42489\pm0.00011$ days.
Thus, if we found degenerate solutions to equation \eqref{eq:t_T_2} and \eqref{eq:t_T_3} we could check whether one of these is larger than $P_{\text{max}}$ and rule it out, even before resolving equation \eqref{eq:complex_eq}. 
We note that equations \eqref{eq:simple_eq} and \eqref{upper_period} differ from equations (27)-(28) of S03 because of our different definition of the mean stellar density. 

All the equations presented in this work are summarised in Table \ref{tab:summary_table} in Appendix \ref{app:summary}, together with the assumptions under which they can be applied.

\subsection{Beyond the circular orbit approximation}
\label{sec:beyond_circular}
Our framework highly relies on the circular orbit approximation but exoplanets may not have a null eccentricity. The orbital eccentricity affects the shape of a planetary transit (e.g., \citealt{Winn2007}). \cite{Tingley2005} were the firsts to examine the influence of the planetary eccentricity on the transit light curve. 
Furthermore, \cite{Barnes2007} demonstrated that if the star radius is constrained by an independent measurement, transit photometry can provide a lower limit on planetary eccentricity.

Planets on eccentric orbits have a higher transit probability than planets in circular orbit with the same semi-major axis \citep{Barnes2007}. Moreover, transit events for exoplanets on eccentric orbits may exhibit asymmetric shapes or varying durations compared to those on circular orbits \citep{YeeGaudi2008}. As a result, the duration of the transit depends on where the planet is in its orbit.
Thus, for eccentric planets, equation \eqref{eq:t_T} involves the planetary eccentricity $e$ and the argument of periastron $\omega$ and becomes
\begin{equation}
    t_T=\dfrac{P}{\pi}\arcsin{\left(\dfrac{R_*}{a}\left[\dfrac{[1+\sqrt{\delta}]^2-[a\cos i/R_*]^2}{1-\cos ^2 i}\right]^{1/2}\right)}\dfrac{\sqrt{1-e^2}}{1+e\sin\omega}.
    \label{eq:t_T_ecc}
\end{equation}
If we define
\begin{equation}
    \tau\equiv t_T\left(\dfrac{1+e\sin\omega}{\sqrt{1-e^2}}\right),
\end{equation}
then equation \eqref{eq:t_T_ecc} can be written as
\begin{equation}
    \tau=\dfrac{P}{\pi}\arcsin{\left(\dfrac{R_*}{a}\left[\dfrac{[1+\sqrt{\delta}]^2-[a\cos i/R_*]^2}{1-\cos ^2 i}\right]^{1/2}\right)},
    \label{eq:t_T_ecc2}
\end{equation}
which is structurally identical to equation \eqref{eq:t_T}. Thus equation \eqref{eq:t_T_ecc2} admits the solutions $P_1$ and $P_2$ given by equations \eqref{eq:P1} and \eqref{eq:P2} where $t_T\rightarrow\tau$.

\begin{figure*}
    \centering
    \includegraphics[width=\textwidth]{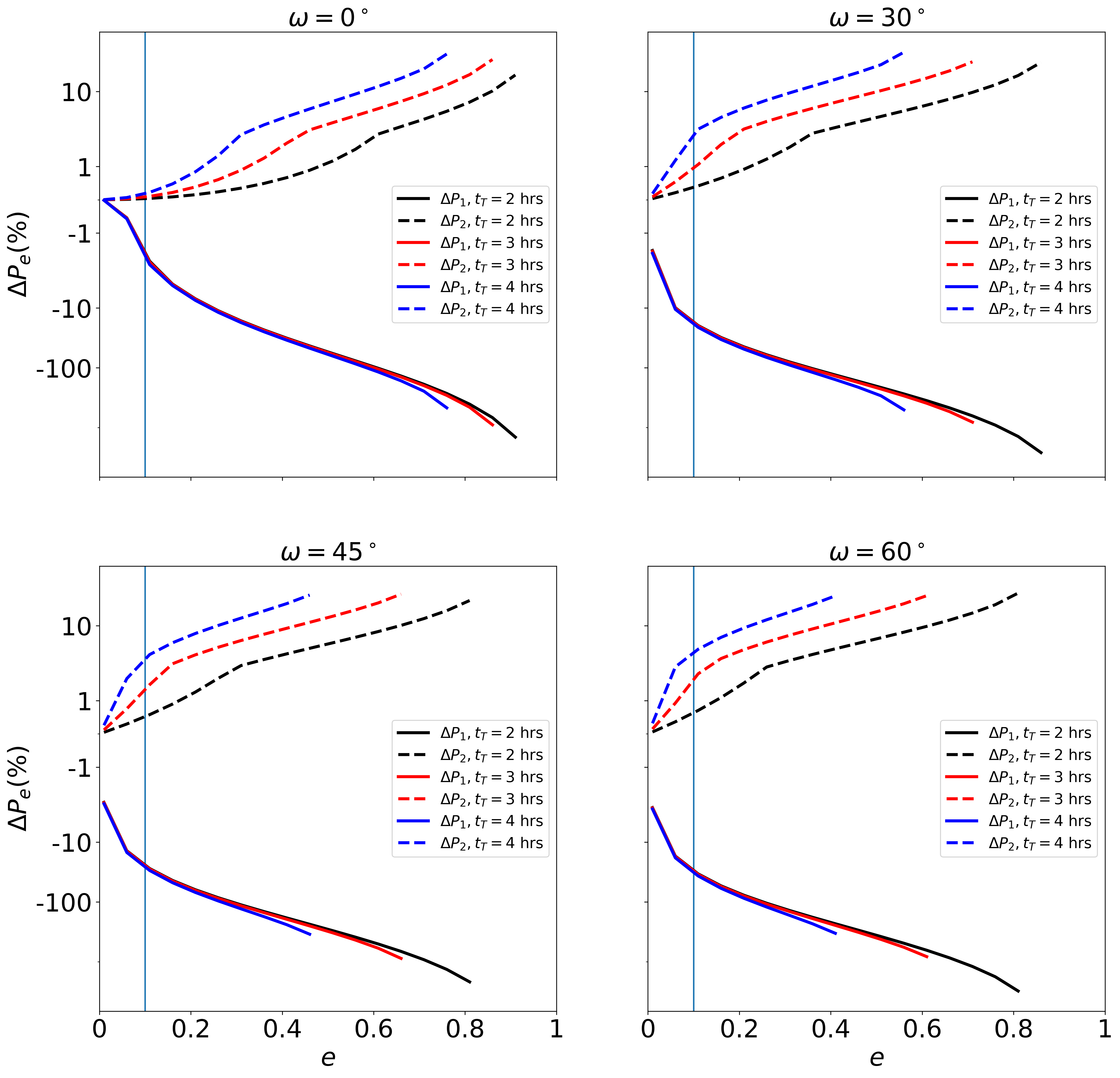}
    \caption{The solutions $P_1$ (solid lines) and $P_2$ (dashed lines) as function of the planetary eccentricity $e$ and argument of periastron $\omega$ for different values of the transit duration $t_T$ (black $2$ hours, red $3$ hours and blue $4$ hours). The quantity $\Delta P_e\equiv (P_e-P_0)/P_0$, where $P_e$ is the orbital period for non-zero eccentricity and $P_0$ corresponds to the orbital period if the planet is on a circular orbit. Each panel refers to a different value of the argument of periastron $\omega$. In each panel the vertical blue line indicates $e=0.1$.}
    \label{fig:ecc_behav}
\end{figure*}

In Fig. \ref{fig:ecc_behav} we show how the $P_1$ and $P_2$ solutions depend on the planetary eccentricity $e$ for different values of the argument of periastron $\omega$ and transit duration $t_T$, once  the orbital inclination $i$, the stellar density $\rho_*$ and the transit depth $\delta$ are fixed. Firstly, we observed that the $P_2$ solution in the circular orbit approximation is systematically underestimated with respect to that for non-zero planetary eccentricity, and viceversa for the $P_1$ solution. Furthermore, we found that the second solution, $P_2$, is more stable than $P_1$ under the circular orbit approximation. While the former can vary by up to $\approx 20\%$ from its value in the circular orbit approximation, the latter can change by more than $100\%$ from its prior value. Thus, if the evaluation of the orbital period using equation \eqref{eq:simple_eq} alongside its uncertainty rules out the $P_1$ solution, then our prediction is based on $P_2$. In this case we expect a discrepancy of up to $20\%$ in case of non-zero eccentricity. Conversely, the most problematic scenario is when the retrieved orbital period comes from the $P_1$ solution or, if the uncertainty $\Delta P_{t_F}$ over $P_{t_F}$ is large enough, a combination of $P_1$ and $P_2$.

When the planet orbits on a mild eccentric orbit (i.e. $e<0.1$) both solutions vary by less than $30\%$ from the circular orbit case. To date $\approx 50\%$\footnote{https://exoplanetarchive.ipac.caltech.edu/} of exoplanets with known eccentricity have $e<0.1$.  Furthermore, once the argument of periastron $\omega$ is fixed, increasing the transit duration $t_T$ leads to a narrower range of potential eccentricities and a more pronounced deviation of the $P_2$ solution from the circular orbit case.

Lastly, we also analysed the behaviour of $P_1$ and $P_2$ as function of the argument of periastron $\omega$. In particular, since $\tau\propto\sin\omega$, it is sufficient to restrict our analysis in the ranges $0^\circ \leq \omega \leq 90^\circ$ and $180^\circ\leq\omega\leq 270^\circ$. We found that when $\omega=0^\circ$ increases to $\omega=90^\circ$ the range of possible eccentricities diminishes while the deviation from the circular case increases for both $P_1$ and $P_2$. Conversely we did not find appreciable differences when $\omega=180^\circ$ increases to $\omega=270^\circ$.

A non-zero eccentricity not only has a significant impact on the transit parameters, but it may also result in false positive scenarios that are not geometrically feasible under the circular orbit approximation as we will discuss in Section \ref{sec:vetting}. 

\subsection{Comparative analysis}
\label{sec:test}
\begin{figure*}
    \centering
    \includegraphics[width=\textwidth]{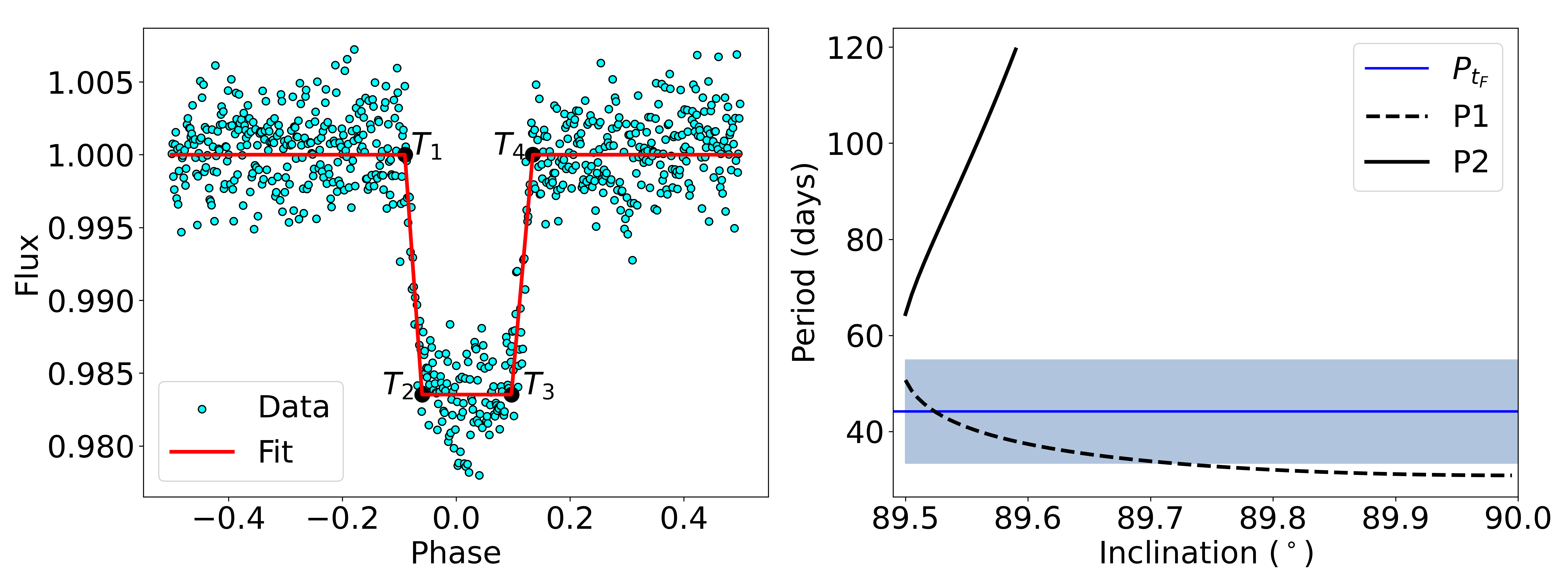}
    \caption{Results of our method applied to the giant warm Jupiter TOI 216.02. \textit{Left:} the TOI 216.02 transit in Sector 1 alongside the best-fitting trapezoid model. This model allows us to estimate $T_1 , T_2 , T_3$ and $T_4$ which respectively correspond to the first, second, third and four contact in the no-limb darkening assumption. The total transit duration is $t_T=T_4-T_1$ while the elapsed time from the second to the third contact $t_F=T_3-T_2$. \textit{Right:} the solid and dashed black lines represent the solutions $P_2$ and $P_1$ respectively as function of the orbital inclination $i$. There is a gap between the two functions between around $57$ and $65$ days. The blue solid line represents the value of $P_{t_F}$ and the blue shaded area is its correspondent uncertainty. The range of possible periods, along with a spectrum of inclinations, is defined by the intersection between the blue shaded area and $P(i)$.}
    \label{fig:toi216.02}
\end{figure*}

To evaluate the effectiveness of the procedure outlined in Sections \ref{sec:polynomial_eq}, \ref{sec:uncertainties}, and \ref{sec:degeneracy}, we here show its application to TOI 216.02, a giant warm Jupiter ($R_P=8.0^{+3.0}_{-2.0}\, R_\oplus$) orbiting its host star in $34.556_{-0.010}^{+0.014}$ days on a $89.89_{-0.12}^{+0.11}$ degree inclined and mild eccentric ($e=0.160^{+0.003}_{-0.002}$) orbit \citep{Kipping2019}. It has been detected by TESS for the first time in Sector 1 at $\approx 1331.26$ BTJD. We then used this single transit measurement to perform a best-fitting trapezoid model as shown in Fig. \ref{fig:toi216.02}. The best-trapezoid transit fit allows us to estimate $t_T, t_F$ and $\delta$. We retrieved the mean stellar density $\rho_*$ from the Gaia DR3 archive \citep{GaiaCollaboration2016, GaiaCollaboration2022}. 
Next, we calculated $P_1$ and $P_2$ (from equations \eqref{eq:P1}-\eqref{eq:P2}) for each inclination value given by the inequality \eqref{eq:grid}. If either $P_1$ or $P_2$ exceeded the threshold $P_{\text{max}}$, we discarded them as viable solutions. Ultimately, our procedure yielded the functions $P_1(i)$ and $P_2(i)$. 
Fig. \ref{fig:toi216.02} depicts the outcomes of our method when applied to TOI 216.02. We observe that both $P_1(i)$ and $P_2(i)$ branches cover a broad range of periods, extending from approximately $32$ days to around $120$ days, corresponding to orbital inclinations in the range of $89.5^\circ$ to $90.0^\circ$. The issue of degeneracy becomes apparent, resulting in two potential orbital periods for a given inclination. To break the degeneracy and put constraints on both $P$ and $i$ at the same time, we computed a preliminary estimate of the orbital period using equation \eqref{eq:simple_eq} obtaining $P_{t_F}=44\pm 11$ days. 
This strongly constrains the possible orbital periods and inclinations. As shown in Fig. \ref{fig:toi216.02}, the $P_{t_F}$ estimate and its related uncertainty intersect the $P_1$ branch breaking the $P_1-P_2$ degeneracy. The uncertainty $\Delta P_{t_F}$ defines a range of possible values for the orbital period and inclination. We take the median of their respective ranges as the best estimate value for $P$ and $i$, while their uncertainties correspond to the semi-difference of this range. Thus, the orbital period and inclination ranges are narrower as $\Delta P_{t_F}$ decreases. Furthermore, from the best-trapezoid fit, we also obtain an estimate of the planetary radius which is $R_p=11.22\pm 0.81\, R_\oplus$.
In summary, our approach yields an orbital period of $P = 42.1 \pm 8.6$ days and an orbital inclination of $i=89.6\pm 0.1^\circ$. We note that our overestimation by $\approx 20\%$ from the actual orbital period is expected since the solution $P_1$ is more sensitive to non-zero eccentric orbit as discussed in Section \ref{sec:beyond_circular}.
We here stress that the precision of our approach strongly depends on the quality of the transit fit. Specifically, the precision of parameters $t_T$, $t_F$, and $\delta$ directly impacts the uncertainty associated with $P_{t_F}$ as discussed in Appendix \ref{appendix: formal_solutions}. 
A more precise determination of these parameters will result in a smaller uncertainty on $P_{t_F}$ which in turn leads to a narrower range of orbital periods and inclinations.

\begin{figure*}
    \centering
    \includegraphics[width=\textwidth]{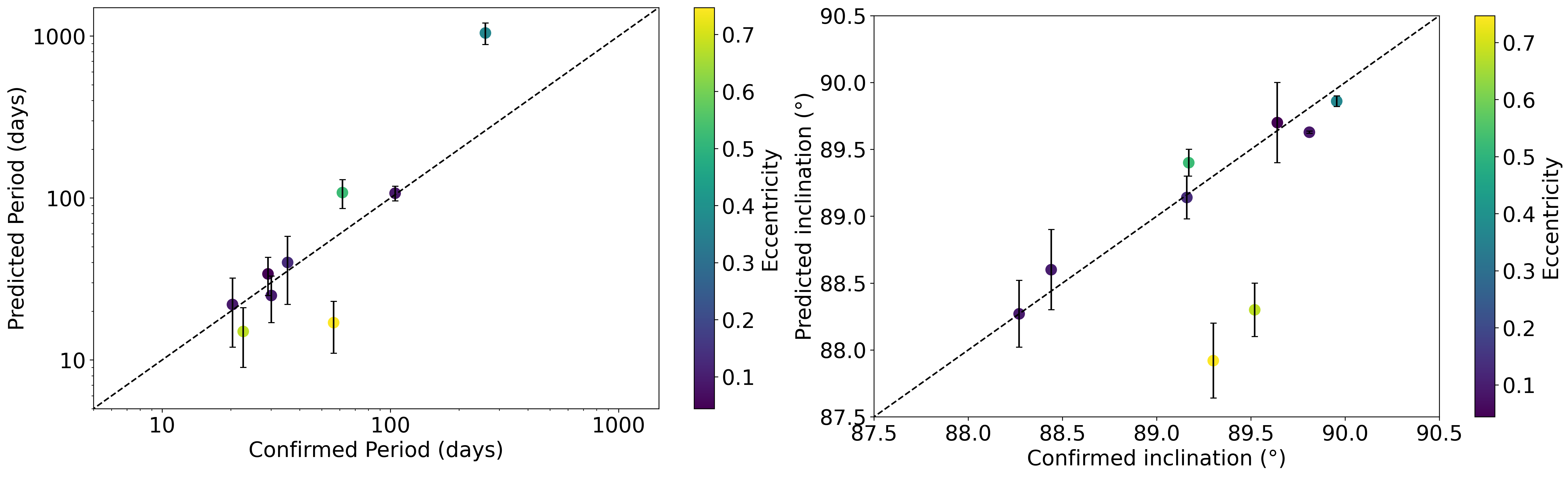}
    \caption{Comparative analysis of our method applied on mono-transit TESS signals confirmed as true planets by follow-up observations. For each planet of the sample test, in the \textit{left} panel we show our predicted period against its true value, while in the \textit{right} panel we depict our predicted orbital inclination against its true value. Each planet is colored according to its eccentricity. In both plots, the black dashed line represents the agreement between our prediction and the literature. We remark that our method mainly fails when the circular orbit assumption is not satisfied.}
    \label{fig:test_comparison}
\end{figure*}

Subsequently, we applied our methodology to several mono-transit events identified in the TESS data that were later confirmed as genuine planets through subsequent observations. The outcomes of our comparative analysis are presented in Table \ref{tab:comp_analysis} and shown in Fig. \ref{fig:test_comparison}. 
There is a good agreement between our method and the results from the literature when the assumption of a circular orbit is met, as seen in the cases of TOI 199 b, TOI 1847 b, TOI 1899 b, TOI 4406 b and TOI 5153 b. However, the method fails to retrieve the orbital parameters in cases where the assumption of a circular orbit is not satisfied. This was expected since the circular orbit assumption is the first and the most relevant hypothesis at the core of our framework. We remark the only exception in our sample test is TOI 2338 b for which we found a good agreement despite its high-eccentric orbit. Furthermore, we observed that a simple trapezoid fit was inadequate for cases with asymmetric transits caused by high-eccentricity orbits, as exemplified by TOI 5152 b.

\begin{table*}
\begin{tabular}{ccccc}
\hline
TOI    & Period (days)            & $i (^\circ)$               & $e$                                                                          & Source                                    \\ \hline
199 b & $104.87236\pm 0.00005$    & $89.81\pm 0.02$    & $0.09^{+0.01}_{-0.02}$                                                      & \citealt{Hobson2023}       \\\
       & $107\pm 11$               & $89.63 \pm 0.01$           & -                                                                            & This work                                 \\ \hline
1847 b & $35.45533\pm 0.00019$    & $89.16^{+0.20}_{-0.29}$    & $0.13^{+0.10}_{-0.09}$                                                       & \citealt{Gill2020}       \\\
       & $40\pm 18$               & $89.14 \pm 0.16$           & -                                                                            & This work                                 \\ \hline
1899 b & $29.090312^{+0.000036}_{-0.000036}$    & $89.64^{+0.07}_{-0.08}$    & $0.044^{+0.029}_{-0.027}$                                   & \citealt{Lin2023}       \\\
       & $34\pm 9$               & $89.7 \pm 0.3$           & -                                                                            & This work                                 \\ \hline
2180 b & $260.79^{+0.59}_{-0.58}$ & $89.955^{+0.032}_{-0.044}$ & $0.3683\pm 0.0073$                                                           & \citealt{Dalba2022}      \\\
       & $1044 \pm 158$           & $89.86 \pm 0.04$           & -                                                                             & This work                                 \\ \hline      
2338 b & $22.65398\pm 0.00002$ & $89.52\pm 0.02$ & $0.676\pm 0.002$                                                           & \citealt{Brahm2023}      \\\
       & $15\pm 6$      & $88.3\pm 0.2$           & -                                                                             & This work                                 \\ \hline      
2589 b & $61.6277\pm 0.00002$ & $89.17\pm 0.01$ & $0.522\pm 0.006$                                                           & \citealt{Brahm2023}      \\\
       & $108\pm 22$      & $89.4\pm 0.1$           & -                                                                             & This work                                 \\ \hline      
4127 b & $56.39879\pm 0.00010$    & $89.30^{+0.46}_{-0.60}$              & $0.7471^{+0.0078}_{-0.0086}$& \citealt{Gupta2023} \\\
       & $17\pm 6$                       & $87.92\pm 0.28$                          & -                                                                            & This work                                 \\ \hline
4406 b & $30.08364\pm {+0.00005}$ & $88.44\pm 0.04$ & $0.1^{+0.04}_{-0.05}$                                                           & \citealt{Brahm2023}      \\\
       & $25 \pm 8$           & $88.6 \pm 0.3$           & -                                                                             & This work                                 \\ \hline 
5152 b & $54.18915\pm 0.00015$    & $88.4\pm 0.6$              & $0.432\pm 0.023$& \citealt{Ulmer-Moll2022} \\\
       & -                        & -                          & -                                                                            & This work                                 \\ \hline
5153 b & $20.33003\pm 0.00007$    & $88.27\pm 0.14$            & $0.0910^{+0.0240}_{-0.0260}$                                                 & \citealt{Ulmer-Moll2022} \\\
       & $22 \pm 10$              & $88.27 \pm 0.25$           & -                                                                            & This work                                 \\ \hline
\end{tabular}
\caption{Comparison between the results of our method and values from literature on mono-transit TESS events later confirmed as true planets by follow-up observations.}
\label{tab:comp_analysis}
\end{table*}

\section{The sample}
\label{sec:sample}
Cross-matching mono-transit TESS candidates with the all-sky PLATO Input Catalog (asPIC; \citealt{Montalto2021}) can aid in the planning and prioritization of follow-up observations. If a mono-transit candidate is associated with an asPIC entry, it indicates the potential future availability of additional high-precision photometry and asteroseismic data, should a LOP (long-observing phase; at least two years of continuous coverage) or a SOP field (short-observing phase; at least two months) be overlapped with it. This additional information can guide the selection of suitable candidates for further ground-based observations, such as radial velocity measurements or atmospheric characterization, increasing the chances of confirming and studying the exoplanet.

\begin{figure*}
    \centering
    \includegraphics[width=0.85\textwidth]{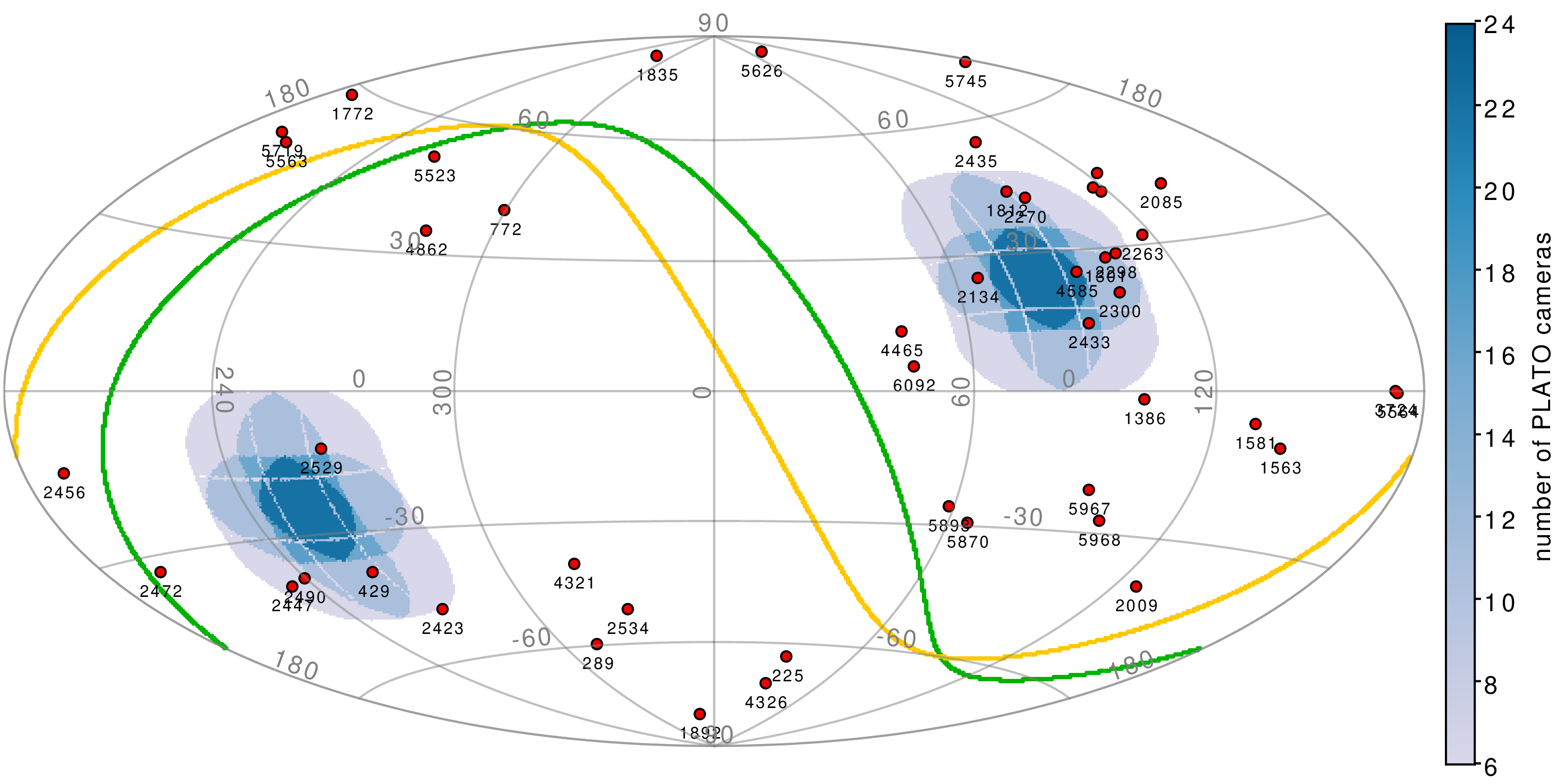}
    \caption{All-sky Aitoff projection in Galactic coordinates, showing the footprints of the two LOP fields (LOPS2 and LOPN1; color-coded in blue shades according to the number of co-pointing cameras), the mono-transit targets identified in this work (red points with the TOI IDs as labels), the Ecliptic (yellow line) and the celestial equator (green line). A zoom-in version centered on the two LOP fields is given in Fig.~\ref{fig:lop_plato_zoom}.}
    \label{fig:lop_plato}
\end{figure*}

As of May 2023, TESS detected $83$ mono-transit events whose key features are publicly available in the ExoFOP archive. Out of these $83$ events, $45$ have been detected by the Science Processing Operations Center pipeline (SPOC, \citealt{Jenkins2016}), $24$ by the Quick-look Pipeline (QLP, \citealt{Huanga, Huangb}) and the last $14$ are Community TOIs \citep{Guerrero2021}. TOI 289.01 is the oldest mono-transit event of the sample to be detected (3rd October 2018), while the latest candidates have been detected on the 2nd February 2023.
However, $23$ of these signals have been ruled out as false positives or false alarms by the TESS Follow-up Observing Program Working Group (TFOPWG). We decided to remove them from our analysis to avoid duplicated work. We also filtered out all the transit events with $S/N<7.3$, which is a standard threshold value used to refer to a planet as detectable \citep[see e.g.][]{Sullivan2015,Bouma2017,Barclay2018,Villanueva2019}. 
Thus, we selected 57 TESS mono-transit planet candidates with $S/N\geq 7.3$. Then, we cross-matched this sample with the asPIC \citep{Montalto2021} to focus on those mono-transit TESS candidates orbiting a star that could also be monitored by PLATO in the future. After performing the cross-matching using \texttt{topcat} we are left with a catalog of $48$ mono-transit events detected by TESS that are included in the asPIC. They are plotted as labelled red circles in the sky map of Fig.~\ref{fig:lop_plato} and listed in Table~\ref{tab:monotransits_candidates}.

\subsection{PLATO LOP fields}
\begin{figure*}
    \centering
    \includegraphics[width=0.5\textwidth]{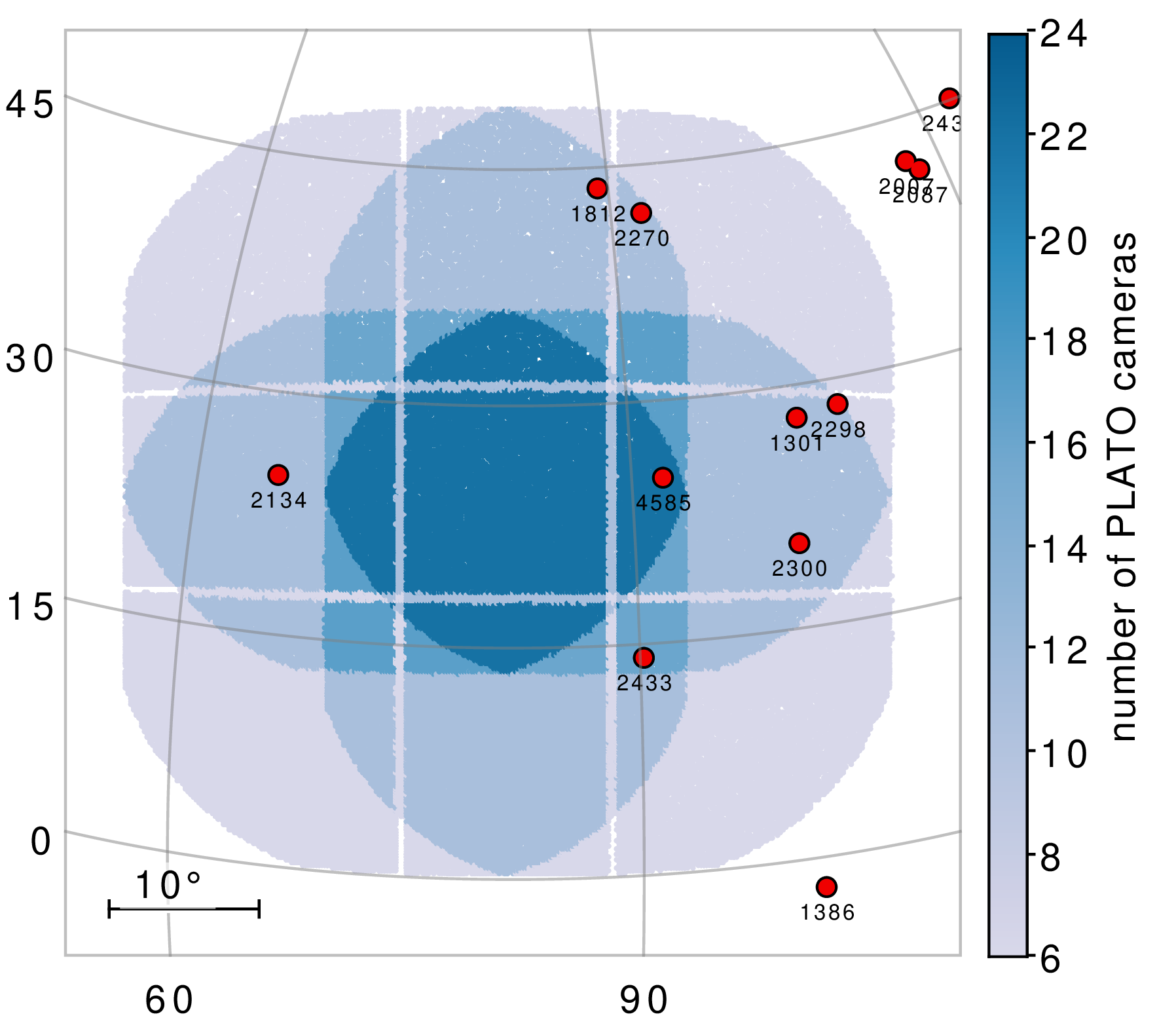}\includegraphics[width=0.5\textwidth]{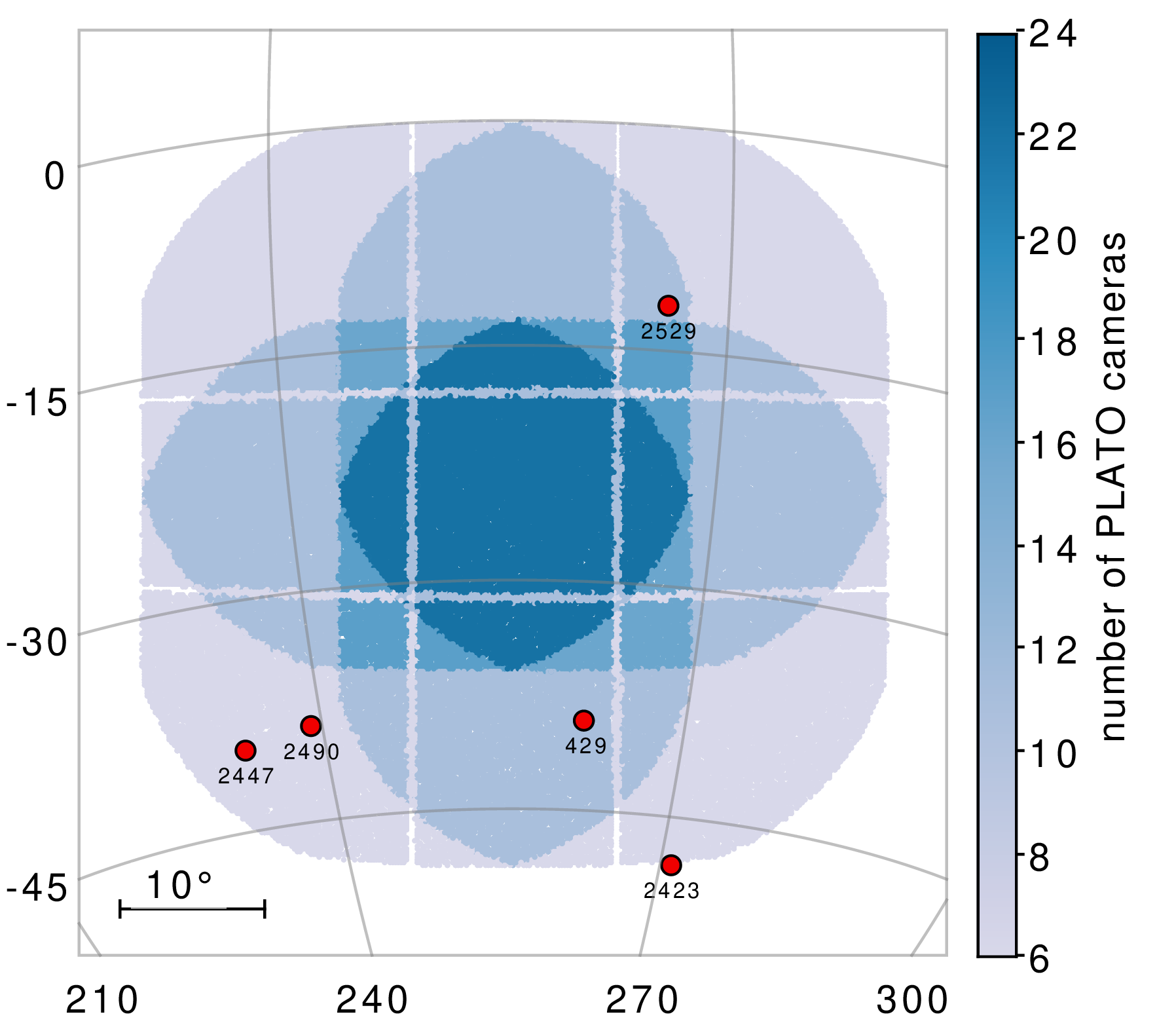}
    \caption{As in Fig.~\ref{fig:lop_plato}, but zoomed in on the PLATO long-duration fields LOPN1 (left panel) and LOPS2 (right panel). The latter is the LOP field formally selected as the first one to be pointed by PLATO in 2027. Projection is orthographic in Galactic coordinates.}
    \label{fig:lop_plato_zoom}
\end{figure*}

It has been already formally announced that PLATO will start its nominal mission at the end of 2026 by pointing a LOP field in the Southern hemisphere for at least two continuous years. This field, named LOPS2, is a slightly modified version of the LOPS1 field identified by \citet{Nascimbeni2022} and will be described in detail in a forthcoming paper (Nascimbeni et al., in prep.). Four mono-transit candidates lie within the LOPS2 boundaries (Fig.~\ref{fig:lop_plato_zoom}, right plot): TOI-429 and TOI-2529 in the 12-telescope regions, TOI-2447 and TOI-2490 in the six-telescope regions. A fifth candidate, TOI-2423, is located within a few arcminutes from the LOPS2 outer rim, and could be possibly fall on silicon depending on the actual pointing and co-alignment perfomances of PLATO.

A second, northern LOP field (LOPN1) has been identified as well by \citet{Nascimbeni2022}. While it is not guaranteed that it will be observed, we did the same excercise as above, by identifying eight mono-transit targets within its boundaries (Fig.~\ref{fig:lop_plato_zoom}, left plot). One of those, TOI-4585, lies in the 24-telescope region, i.e., where the field of view of all the PLATO normal cameras (NCAMs) is overlapped.

\subsection{Vetting}
\label{sec:vetting}
The next step in our analysis consists in uniformly vetting the $48$ mono-transit events in order to produce a clean catalog. 
Vetting is a crucial step in the process of confirming the nature of candidate exoplanets. It involves a rigorous analysis of various observational and statistical factors to distinguish genuine exoplanets from false positives, which are astrophysical signals (eclipsing binary stars, stellar spots etc.) or instrumental artefacts (e.g., jitter noise and momentum dumps) that mimic the signals of transiting exoplanets \citep{Ciardi2018}. 
Planet candidates are generally vetted firstly by an-eye inspection at both the light curve and the pixel level (see for example \citealt{Magliano2023Neptunes,Magliano2023Patrol}); then if the candidate passes this first investigation, vetters try to statistically validate it calculating the false positive probability by means of tools like \texttt{VESPA} \citep{Morton2012,Morton2015} and \texttt{TRICERATOPS} \citep{Giacalone2021}. 
Following this procedure for a mono-transit planet candidate poses inherent challenges due to limited available data and the associated uncertainties. Insufficient observations and data gaps hinder a comprehensive understanding of these events, making it difficult to confidently ascertain the true nature of mono-transit signals. 
Hence, it is only more likely for a single event in a photometric survey to be a false alarm compared to a periodic signal. 
Moreover, a single measurement may not provide robust statistical evidence to classify that event as a false positive. 
For this reason, in our vetting analysis, we opted for an extremely cautious approach considering signals as potential planet candidates even if they exhibited minor flags (e.g., low signal-to-noise ratio detections that can be challenging to interpret, particularly based on a single measurement alone).
Nevertheless, it is still possible to rule out a candidate if a clear systematic in the field-of-view (FoV) occurs at the time of the transit. The presence of a sudden flux change in the background, occurring at or close to the transit event, can introduce false signals into the light curve, effectively imitating or distorting the transit signal. Moreover, we also looked for any centroid offset (CO) which hints the transit could have originated from a nearby source that falls within the aperture mask used to extract the target's light curve. 
Therefore, we thoroughly examined both the light curve and the surrounding background flux for each candidate with a custom pipeline alongside the LATTE web interface \footnote{http://latte-online.flatironinstitute.org/app}\citep{Eisner2020}. Any anomalies, such as atypical characteristics or discontinuities in the background flux, prompt us to flag the transit with a potential issue.
Lastly, as anticipated in Section \ref{sec:beyond_circular}, under certain geometrical configurations ($e>0, 180^\circ\leq\omega\leq 360^\circ$ and $i\neq 90^\circ$) an eclipsing binary could present only the secondary eclipse \citep{Santerne2013} while the primary falls outside the line of sight. If the secondary eclipse is shallow enough, it can mimic a planetary transit leading to a false positive detection (e.g., KOI-419 and KOI-698 binary system discussed in \citealt{Santerne2012}). \cite{Santerne2013} showed that secondary-only eclipsing binary systems produce a V-shaped transit profile which however could be also caused by a grazing transiting giant planet. This devious false positive scenario can be ruled out if a radial velocity follow-up measurement finds a strong variation in anti-phase with the photometric ephemeris. Since in this work we are just limiting our analysis to the light curve of the star, we cannot inspect this possibility but only flag whether the transit exhibits a V-shaped feature.

At the end of the vetting process we passed \npcs signals as Planet Candidate (PC) while ruling out \nfps events as False Positive (FP). For each signal, we also checked whether any other sources fall within the same pixel of the target by consulting stellar catalogs \citep{Wenger2000, GaiaCollaboration2022}. In this case, we flagged the candidate with the acronym FSCP (Field Star in Central Pixel). This is a common issue when dealing with TESS due to its low angular resolution ($1\text{px}\approx 21''$) which imposes an inherent limitation on the vetting process. Blended sources falling within the same pixel can not only contaminate the starlight, resulting in a shallower transit but can also cause the observed transit. However, this flag alone is not sufficient to rule out a transiting-like event as a FP. Nevertheless, it is important to keep this in mind because, as mentioned in Sect. \ref{sec:methods}, our method relies on the assumption of no contamination by other sources. Thus, a FSCP flag indicates the possibility the orbital period estimates could be revised using a less contaminated light curve \citep{Han2023}. Table \ref{tab:monotransits_candidates} presents a comprehensive list of the $48$ TESS mono-transit candidates analyzed in this study, including their disposition and corresponding comments following our thorough vetting analysis. After removing the false positive signals, our final dataset contains \npcs good mono-transit TESS candidates. 

\begin{table*}
\begin{threeparttable}
\begin{tabular}{lll|lll|lll}
\hline
TOI     & Disposition & Comments                      & TOI     & Disposition & Comments                      & TOI     & Disposition & Comments                           \\ \hline
225.01  & FP          & Background event              & 2134.02 & PC          & High SNR                      & 4321.01 & PC          & LowSNR       \\
289.01  & PC          & High SNR                      & 2263.01 & PC          & FSCP, lowSNR                  & 4326.01 & PC          & LowSNR        \\
429.01  & PC          &                               & 2270.01 & FP          & NT in SAP light curve & 4465.01 & PC          & {}           
\\
772.02  & PC          & Multiple system               & 2298.01 & PC          & FSCP, lowSNR                  & 4585.01 & PC          & FSCP, low SNR \\
1301.02 & FP          & NT\tnote{†}\, in SAP light curve & 2300.03 & PC          & High SNR, Long duration       & 4862.01 & PC          & FSCP         \\
1386.01 & PC          & FSCP                          & 2423.01 & PC          &                               & 5523.01 & PC          & Long duration \\
1563.01 & PC          & FSCP                          & 2433.01 & PC          & FSCP, low SNR                 & 5563.01 & PC          &               \\
1581.01 & FP          & V-shaped transit              & 2435.01 & FP          & NT in SAP light curve & 5564.01 & PC          & FSCP        
\\
1772.02 & PC          & High SNR                      & 2436.01 & PC          & Low SNR                       & 5626.01 & PC          &               \\
1812.01 & PC          & High SNR                      & 2447.01 & PC          & High SNR                      & 5719.01 & PC          & LowSNR        \\
1835.02 & PC          & Low SNR                       & 2456.01 & FP          & Background event              & 5745.01 & PC          &               \\
1892.01 & PC          & Low SNR,FSCP                  & 2472.01 & PC          & Low SNR                       & 5870.01 & PC          &               \\
2007.01 & FP          & CO                           & 2490.01 & PC          & High SNR                      & 5893.01 & PC          &               \\
2009.01 & FP          & FSCP, V-shaped transit        & 2529.01 & PC          & High SNR                      & 5967.01 & PC          & LowSNR       \\
2085.01 & PC          & FSCP                          & 2534.01 & PC          & FSCP                          & 5968.01 & PC          & LowSNR        \\
2087.01 & FP          & NT in SAP light curve & 3724.01 & PC          &                               & 6092.01 & FP          & FSCP, CO      \\ \hline
\end{tabular}
\begin{tablenotes}
\item[†] No Transit. 
\end{tablenotes}
\end{threeparttable}
\caption{Summary vetting report of the 48 monotransit TESS planet candidates orbiting stars that will be monitored by PLATO. We ruled out \nfps of them as false positives (FP) while keeping \npcs as planet candidate (PC). The acronyms CO and FSCP stand for "Centroid Offset" and "Field Star in Central Pixel" respectively.}
\label{tab:monotransits_candidates}
\end{table*}

\subsection{Stellar parameters}
We obtained the stellar properties of the \npcs host stars by utilizing \texttt{ARIADNE} \citep{Vines2022} which models the stellar Spectral Energy Distribution (SED). \texttt{ARIADNE} integrates six atmospheric model grids which have been previously convolved with various filters' across multiple bandpasses. The utilized models include \texttt{Phoenix v2} \citep{Husser2013}, BT-Settl, BT-Cond \citep{Allard2011}, BT-NextGen \citep{Hauschildt1999,Allard2011}, \cite{Kurucz1993} and \cite{Castelli2003}. \texttt{ARIADNE} uses the (Dynamic) Nested Sampling algorithm \citep{Skilling2004,Skilling2006,Higson2019} via \texttt{dynesty} \citep{Speagle2019} for each of the aforementioned models. It conducts Bayesian Model Averaging over all fitted models to derive the final set of parameters for a given star. Initially, \texttt{ARIADNE} seeks broadband photometry through \texttt{astroquery} \citep{astropy:2013, astropy:2018, astropy:2022}, accessing MAST and VizieR archives to query catalogs such as Tycho-2 \citep{Hog2000}, ASCC \citep{Kharchenko2001}, 2MASS \citep{Skrutskie2006}, GLIMPSE \citep{Churchwell2009}, ALL-WISE \citep{Wright2010}, GALEX \citep{Bianchi2011}, APASS DR9 \citep{Henden2014}, SDSS DR12 \citep{Alam2015}, Strömgren Photometric Catalog \citep{Hauck1998, Paunzen2015}, Pan-STARRS1 \citep{Chambers2016} and Gaia DR2 and DR3 \citep{GaiaDR2, GaiaDR3}. Model selection is based on an initial effective temperature estimation from Gaia DR2, as \cite{Kurucz1993} and \cite{Castelli2003} exhibit poor performance for stars with $T_\text{eff} < 4000 K$. Priors for $T_\text{eff}$ and [Fe/H] are derived from the RAVE catalog \citep{RAVEDR6}, \text{log} $g$ uses a uniform prior ranging from 3.5 to 6. Radius and distance priors follow \texttt{ARIADNE}’s defaults, while interstellar extinction $A_v$ relies on a prior drawn from the Bayestar 3D dustmaps \citep{Green2019} or a uniform prior from $0$ to the maximum line-of-sight extinction from the SFD dustmaps \citep{SFD1,SFD2} when the former is unavailable.

\begin{figure}
    \centering
    \includegraphics[width=0.47\textwidth]{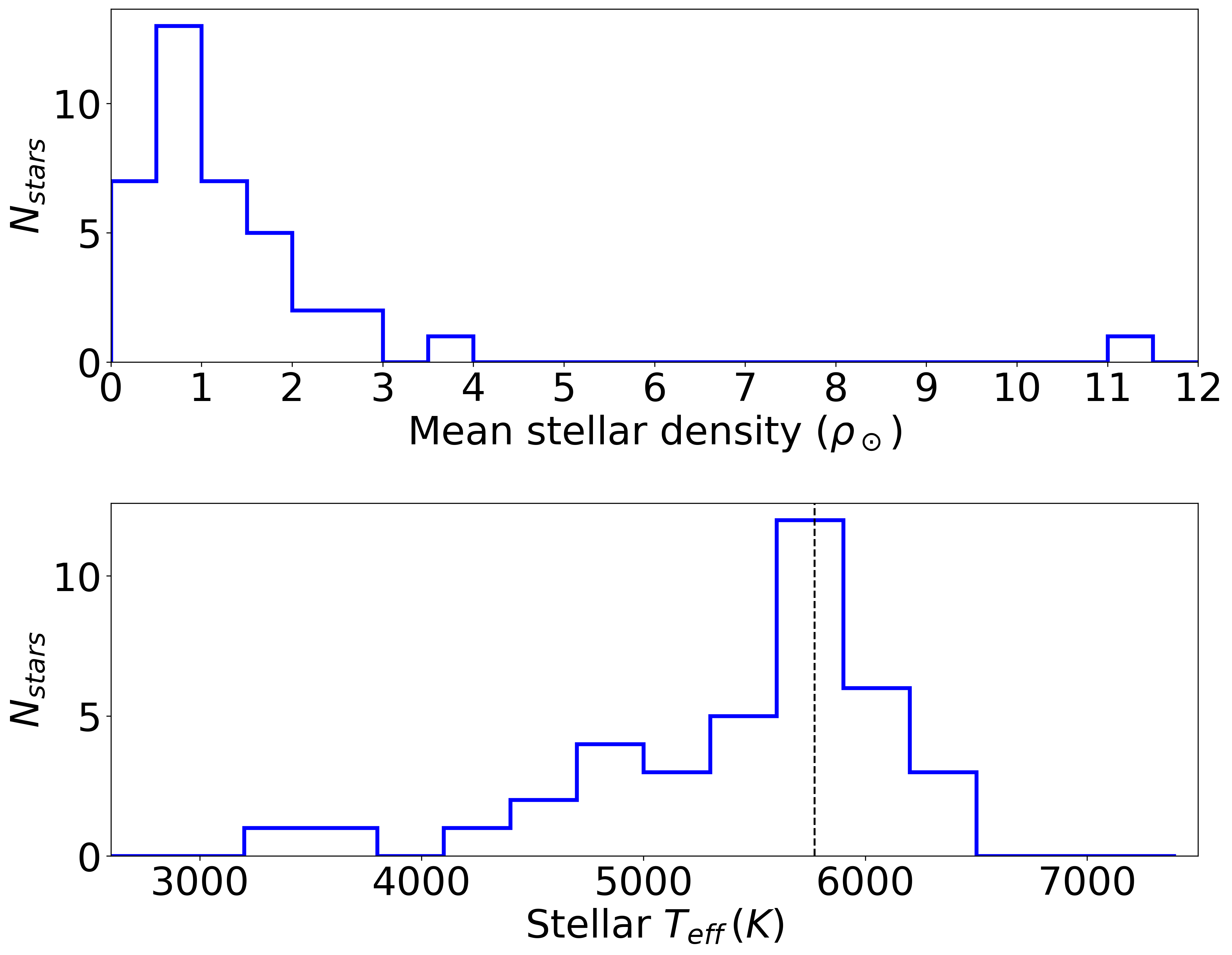}
    \caption{Upper panel: distribution of the mean stellar density $\rho_*$ of the \npcs stars in our sample; lower panel: distribution of the effective temperature of the \npcs stars in our sample. The vertical dashed black line represents the effective temperature of the Sun, i.e. $5772\,K$.}
    \label{fig:distribution_stars}
\end{figure}

In Table \ref{tab:stellar_params} we report the stellar parameters used in this work. In Fig. \ref{fig:distribution_stars} we show the distribution of the \npcs stars within our sample according to their mean stellar density and their temperature as retrieved by \texttt{ARIADNE}. The peak of the two distributions is around $\rho_\odot$ and $T_\odot$ respectively as illustrated in Fig. \ref{fig:distribution_stars}.

\begin{table*}
\begin{threeparttable}
\begin{tabular}{lcccccc}
\hline
TOI & $V_\text{mag}\tnote{†}$ & $T_*(K)$ & $R_* (R_\odot)$ & $\log g$ & $M_*(M_\odot)$ & $\rho_*(\rho_\odot)$ \\ \hline
289   & $ 12.05 \pm 0.32 $ & $ 5676 ^{+ 34 }_{- 53 }$ & $ 1.76 ^{+ 0.015 }_{- 0.034 }$ & $ 3.95 ^{+ 0.10 }_{- 0.08 }$ & $ 1.01 ^{+ 0.04 }_{- 0.06 }$ & $ 0.18 ^{+ 0.01 }_{- 0.02 }$  \\[0.2cm]
429   & $ 10.91 \pm 0.12 $ & $ 5245 ^{+ 25 }_{- 67 }$ & $ 0.946 ^{+ 0.020 }_{- 0.017 }$ & $ 4.48 ^{+ 0.07 }_{- 0.1 }$ & $ 0.99 ^{+ 0.06 }_{- 0.06 }$ & $ 1.17 ^{+ 0.14 }_{- 0.13 }$   \\[0.2cm]
772   & $ 11.68 \pm 0.19 $ & $ 4760 ^{+ 20 }_{- 25 }$ & $ 0.937 ^{+ 0.012 }_{- 0.014 }$ & $ 4.57 ^{+ 0.03 }_{- 0.05 }$ & $ 1.19 ^{+ 0.04 }_{- 0.05 }$ & $ 1.45 ^{+ 0.10 }_{- 0.13 }$   \\[0.2cm]
1386  & $ 10.67 \pm 0.16 $ & $ 5940 ^{+ 115 }_{- 70 }$ & $ 1.023 ^{+ 0.009 }_{- 0.013 }$ & $ 4.44 ^{+ 0.07 }_{- 0.06 }$ & $ 1.05 ^{+ 0.04 }_{- 0.04 }$ & $ 0.98 ^{+ 0.06 }_{- 0.07 }$   \\[0.2cm]
1563  & $ 10.46 \pm 0.15 $ & $ 4662 ^{+ 30}_{-17}$ & $ 0.712 ^{+ 0.010}_{- 0.007}$ & $ 4.60 ^{+ 0.03 }_{- 0.03}$ & $ 0.74^{+0.02}_{-0.02}$ & $2.0  ^{+ 0.1}_{-0.1}$  \\[0.2cm]
1772  & $ 9.96 \pm 0.14 $  & $ 5683 ^{+ 18 }_{- 18 }$ & $ 0.958 ^{+ 0.007 }_{- 0.006 }$ & $ 4.49 ^{+ 0.08 }_{- 0.08 }$ & $ 1.04 ^{+ 0.03 }_{- 0.03 }$ & $ 1.18 ^{+ 0.06 }_{- 0.06 }$  \\[0.2cm]
1812  & $ 12.46 \pm 0.12 $ & $ 4918 ^{+ 31 }_{- 21 }$ & $ 0.758 ^{+ 0.007 }_{- 0.009 }$ & $ 4.58 ^{+ 0.07 }_{- 0.03 }$ & $ 0.8 ^{+ 0.03 }_{- 0.02 }$ & $ 1.84 ^{+ 0.12 }_{- 0.11 }$   \\[0.2cm]
1835  & $ 8.38 \pm 0.12 $  & $ 5310 ^{+ 25 }_{- 26 }$ & $ 0.783 ^{+ 0.009 }_{- 0.008 }$ & $ 4.58 ^{+ 0.04 }_{- 0.04 }$ & $ 0.85 ^{+ 0.03 }_{- 0.02 }$ & $ 1.77 ^{+ 0.12 }_{- 0.10 }$   \\[0.2cm]
1892  & $ 10.61 \pm 0.12 $ & $ 5427 ^{+ 21 }_{- 19 }$ & $ 0.797 ^{+ 0.007 }_{- 0.007 }$ & $ 4.58 ^{+ 0.07 }_{- 0.06 }$ & $ 0.88 ^{+ 0.03 }_{- 0.03 }$ & $ 1.74 ^{+ 0.11 }_{- 0.11 }$   \\[0.2cm]
2085  & $ 10.62 \pm 0.14 $ & $ 5232 ^{+ 67 }_{- 48 }$ & $ 0.792 ^{+ 0.01 }_{- 0.013 }$ & $ 4.57 ^{+ 0.06 }_{- 0.06 }$ & $ 0.85 ^{+ 0.03 }_{- 0.04 }$ & $ 1.71 ^{+ 0.13 }_{- 0.16 }$   \\[0.2cm]
2134  & $ 8.93 \pm 0.15 $  & $ 4483 ^{+ 27 }_{- 31 }$ & $ 0.744 ^{+ 0.012 }_{- 0.012 }$ & $ 4.61 ^{+ 0.03 }_{- 0.04 }$ & $ 0.82 ^{+ 0.03 }_{- 0.03 }$ & $ 1.99 ^{+ 0.17 }_{- 0.17 }$   \\[0.2cm]
2263  & $ 10.34 \pm 0.15 $ & $ 5718 ^{+ 69 }_{- 49 }$ & $ 0.853 ^{+ 0.007 }_{- 0.009 }$ & $ 4.53 ^{+ 0.04 }_{- 0.04 }$ & $ 0.9 ^{+ 0.02 }_{- 0.03 }$ & $ 1.45 ^{+ 0.07 }_{- 0.09 }$   \\[0.2cm]
2298  & $ 12.44 \pm 0.17 $ & $ 4351 ^{+ 18 }_{- 45 }$ & $ 0.629 ^{+ 0.01 }_{- 0.008 }$ & $ 4.67 ^{+ 0.03 }_{- 0.03 }$ & $ 0.68 ^{+ 0.03 }_{- 0.02 }$ & $ 2.73 ^{+ 0.25 }_{- 0.18 }$   \\[0.2cm]
2300  & $ 12.58 \pm 0.25 $ & $ 5577 ^{+ 118 }_{- 94 }$ & $ 0.793 ^{+ 0.009 }_{- 0.011 }$ & $ 4.58 ^{+ 0.04 }_{- 0.03 }$ & $ 0.87 ^{+ 0.03 }_{- 0.03 }$ & $ 1.74 ^{+ 0.12 }_{- 0.13 }$   \\[0.2cm]
2423  & $ 10.04 \pm 0.12 $ & $ 6003 ^{+ 30 }_{- 25 }$ & $ 1.194 ^{+ 0.012 }_{- 0.012 }$ & $ 4.37 ^{+ 0.08 }_{- 0.08 }$ & $ 1.22 ^{+ 0.05 }_{- 0.05 }$ & $ 0.72 ^{+ 0.05 }_{- 0.05 }$   \\[0.2cm]
2433  & $ 14.56 \pm 0.32 $ & $ 3238 ^{+ 51 }_{-31}$ & $ 0.324 ^{+ 0.007 }_{- 0.010 }$ & $ 4.99 ^{+0.13}_{-0.07}$ & $ 0.37 ^{+0.03}_{-0.03}$ & $ 11.0 ^{+1.7}_{-1.7}$  \\[0.2cm]
2436  & $ 9.02 \pm 0.15 $  & $ 6230 ^{+ 37 }_{- 29 }$ & $ 1.238 ^{+ 0.014 }_{- 0.012 }$ & $ 4.3 ^{+ 0.16 }_{- 0.16 }$ & $ 1.12 ^{+ 0.07 }_{- 0.06 }$ & $ 0.59 ^{+ 0.06 }_{- 0.05 }$   \\[0.2cm]
2447  & $ 10.59 \pm 0.12 $ & $ 5821 ^{+ 41 }_{- 32 }$ & $ 1.037 ^{+ 0.013 }_{- 0.013 }$ & $ 4.44 ^{+ 0.10}_{- 0.07 }$ & $ 1.08 ^{+ 0.05 }_{- 0.04 }$ & $ 0.97 ^{+ 0.08 }_{- 0.07 }$   \\[0.2cm]
2472  & $ 9.52 \pm 0.14 $\ & $ 5829 ^{+ 35 }_{- 30 }$ & $ 1.686 ^{+ 0.014 }_{- 0.017 }$ & $ 4.06 ^{+ 0.08 }_{- 0.09 }$ & $ 1.19 ^{+ 0.04 }_{- 0.05 }$ & $ 0.25 ^{+ 0.01 }_{- 0.02 }$   \\[0.2cm]
2490  & $ 11.95 \pm 0.15 $ & $ 5372^{+30}_{-64}$ & $ 1.183 ^{+0.048}_{-0.010}$ & $ 4.37 ^{+0.16}_{-0.05}$ & $ 1.20^{+0.08}_{-0.08}$ & $ 0.7^{+0.1}_{-0.1}$  \\[0.2cm]
2529  & $ 11.4 \pm 0.22 $  & $ 5684 ^{+ 147 }_{- 110 }$ & $ 1.759 ^{+ 0.023 }_{- 0.03 }$ & $ 4.06 ^{+ 0.16 }_{- 0.14 }$ & $ 1.3 ^{+ 0.09 }_{- 0.09 }$ & $ 0.24 ^{+ 0.03 }_{- 0.03 }$   \\[0.2cm]
2534  & $ 11.23 \pm 0.09 $ & $ 6465 ^{+ 33 }_{- 35 }$ & $ 1.616 ^{+ 0.019 }_{- 0.017 }$ & $ 4.11 ^{+ 0.08 }_{- 0.07 }$ & $ 1.23 ^{+ 0.05 }_{- 0.05 }$ & $ 0.29 ^{+ 0.02 }_{- 0.02 }$   \\[0.2cm]
3724  & $ 12.31 \pm 0.60 $ & $ 5927 ^{+ 139 }_{- 94 }$ & $ 1.455 ^{+ 0.019 }_{- 0.018 }$ & $ 4.13 ^{+ 0.06 }_{- 0.07 }$ & $ 1.04 ^{+ 0.04 }_{- 0.04 }$ & $ 0.34 ^{+ 0.03 }_{- 0.03 }$   \\[0.2cm]
4321  & $ 7.78 \pm 0.13 $  & $ 6175 ^{+ 34 }_{- 28 }$ & $ 2.418 ^{+ 0.026 }_{- 0.024 }$ & $ 3.87 ^{+ 0.08 }_{- 0.09 }$ & $ 1.58 ^{+ 0.07 }_{- 0.07 }$ & $ 0.11 ^{+ 0.01 }_{- 0.01 }$   \\[0.2cm]
4326  & $ 8.25 \pm 0.14 $  & $ 5940 ^{+ 29 }_{- 24 }$ & $ 1.123 ^{+ 0.012 }_{- 0.012 }$ & $ 4.3 ^{+ 0.07 }_{- 0.06 }$ & $ 0.92 ^{+ 0.03 }_{- 0.03 }$ & $ 0.65 ^{+ 0.04 }_{- 0.04 }$   \\[0.2cm]
4465  & $ 10.5 \pm 0.17 $  & $ 5783 ^{+ 111 }_{- 77 }$ & $ 0.986 ^{+ 0.009 }_{- 0.012 }$ & $ 4.42 ^{+ 0.05 }_{- 0.05 }$ & $ 0.93 ^{+ 0.03 }_{- 0.03 }$ & $ 0.97 ^{+ 0.06 }_{- 0.07 }$   \\[0.2cm]
4585  & $ 10.22 \pm 0.15 $ & $ 5871 ^{+ 30 }_{- 30 }$ & $ 1.29 ^{+ 0.009 }_{- 0.011 }$ & $ 4.25 ^{+ 0.06 }_{- 0.05 }$ & $ 1.08 ^{+ 0.03 }_{- 0.03 }$ & $ 0.5 ^{+ 0.02 }_{- 0.03 }$   \\[0.2cm]
4862  & $ 12.38 \pm 0.16 $ & $ 5368 ^{+ 47 }_{- 38 }$ & $ 0.915 ^{+ 0.015 }_{- 0.011 }$ & $ 4.48 ^{+ 0.09 }_{- 0.08 }$ & $ 0.92 ^{+ 0.05 }_{- 0.04 }$ & $ 1.2 ^{+ 0.12 }_{- 0.10 }$   \\[0.2cm]
5523  & $ 15.64 \pm 0.33 $ & $ 3679 ^{+ 1 }_{- 33 }$ & $ 0.613 ^{+ 0.001}_{- 0.008 }$ & $ 4.69 ^{+ 0.01 }_{- 0.01}$ & $ 0.67 ^{+ 0.02 }_{- 0.02 }$ & $ 2.91 ^{+ 0.2 }_{- 0.2 }$   \\[0.2cm]
5563  & $ 11.41 \pm 0.15 $ & $ 5130 ^{+ 14 }_{- 68 }$ & $ 1.243 ^{+ 0.065 }_{- 0.002 }$ & $ 4.44 ^{+ 0.11 }_{- 0.01 }$ & $ 1.56 ^{+ 0.20 }_{- 0.01 }$ & $ 0.81 ^{+ 0.23 }_{- 0.01 }$   \\[0.2cm]
5564  & $ 12.36 \pm 0.36 $ & $ 5787 ^{+ 156 }_{- 101 }$ & $ 0.882 ^{+ 0.014 }_{- 0.016 }$ & $ 4.48 ^{+ 0.09 }_{- 0.07 }$ & $ 0.86 ^{+ 0.04 }_{- 0.04 }$ & $ 1.25 ^{+ 0.12 }_{- 0.13 }$   \\[0.2cm]
5626  & $ 10.42 \pm 0.15 $ & $ 5684 ^{+ 32 }_{- 23 }$ & $ 0.977 ^{+ 0.009 }_{- 0.011 }$ & $ 4.46 ^{+ 0.07 }_{- 0.05 }$ & $ 1.01 ^{+ 0.03 }_{- 0.03 }$ & $ 1.08 ^{+ 0.06 }_{- 0.07 }$   \\[0.2cm]
5719  & $ 10.13 \pm 0.14 $ & $ 6296 ^{+ 32 }_{- 31 }$ & $ 1.107 ^{+ 0.012 }_{- 0.011 }$ & $ 4.39 ^{+ 0.09 }_{- 0.10 }$ & $ 1.1 ^{+ 0.05 }_{- 0.05 }$ & $ 0.81 ^{+ 0.06 }_{- 0.06 }$   \\[0.2cm]
5745  & $ 10.62 \pm 0.13 $ & $ 4948 ^{+ 25 }_{- 15 }$ & $ 0.301 ^{+ 0.005 }_{- 0.007 }$ & $ 4.51 ^{+ 0.02 }_{- 0.02 }$ & $ 0.11 ^{+ 0.01 }_{- 0.01 }$ & $ 4.03 ^{+ 0.20 }_{- 0.65 }$   \\[0.2cm]
5870  & $ 11.29 \pm 0.24 $ & $ 5827 ^{+ 40 }_{-7}$ & $ 1.678 ^{+ 0.011 }_{- 0.022 }$ & $ 4.16 ^{+ 0.03}_{- 0.01}$ & $ 1.49 ^{+0.04}_{-0.04}$ & $ 0.31^{+0.02 }_{-0.02}$  \\[0.2cm]
5893  & $ 12.08 \pm 0.22 $ & $ 5853 ^{+ 51 }_{- 79 }$ & $ 1.161 ^{+ 0.015 }_{- 0.022 }$ & $ 4.42 ^{+ 0.07 }_{- 0.09 }$ & $ 1.3 ^{+ 0.05 }_{- 0.08 }$ & $ 0.83 ^{+ 0.06 }_{- 0.10}$   \\[0.2cm]
5967  & $ 10.73 \pm 0.15 $ & $ 4883 ^{+ 55 }_{- 47 }$ & $ 1.103 ^{+ 0.012 }_{- 0.015 }$ & $ 4.38 ^{+ 0.02 }_{- 0.05 }$ & $ 1.07 ^{+ 0.03 }_{- 0.04 }$ & $ 0.8 ^{+ 0.05 }_{- 0.06 }$  \\[0.2cm]
5968  & $ 10.06 \pm 0.13 $ & $ 5997 ^{+ 58 }_{- 56 }$ & $ 1.042 ^{+ 0.011 }_{- 0.009 }$ & $ 4.39 ^{+ 0.06 }_{- 0.06 }$ & $ 0.97 ^{+ 0.03 }_{- 0.03 }$ & $ 0.86 ^{+ 0.05 }_{- 0.05 }$  \\[0.2cm] \hline
\end{tabular}
\begin{tablenotes}
\item[†] Visual apparent magnitude in the Johnson-Cousin system. 
\end{tablenotes}
\end{threeparttable}
\caption{Stellar parameters of the 38 stars hosting mono-transit planet candidates used in this work.}
\label{tab:stellar_params}
\end{table*}
 
\subsection{Notes on false positives}

\begin{figure}
    \centering
    \includegraphics[width=0.47\textwidth]{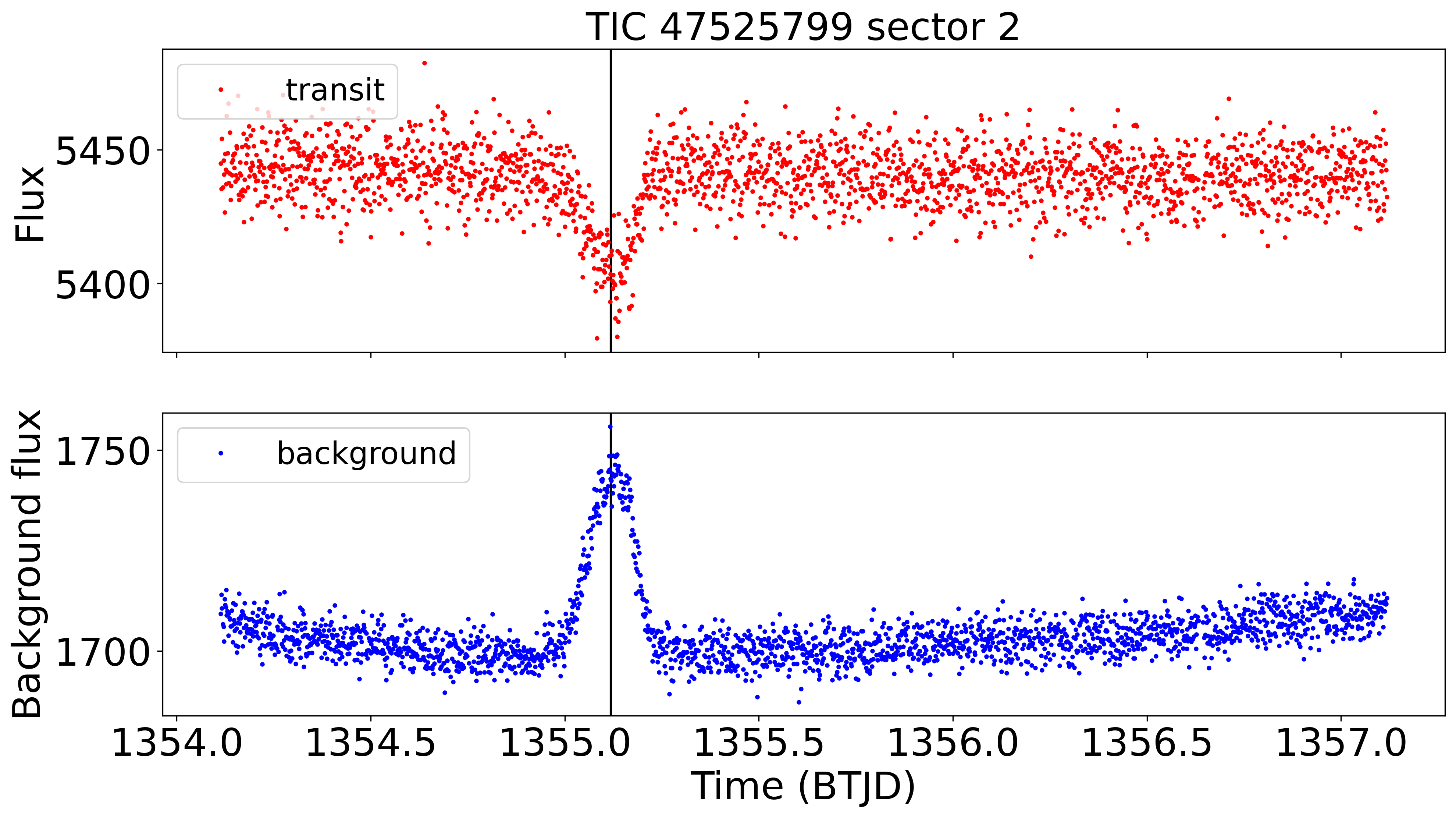}
    \caption{The background flux analysis of planet candidate TOI 225.01 orbiting TOI 225 (or TIC 47525799) observed in sector 2 by TESS. The upper panel presents an enlarged view of the simple aperture photometry flux (depicted in red) during the transit time-frame. The lower panel illustrates the simple aperture photometry background flux (shown in blue) inside the aperture used to calculate the SAP flux, spanning the same time window \citep{Thompson2016,Lightkurve2018}. The vertical black line demarcates the timing of the transit events. We clearly notice a sudden spike in the background flux at the exact time of the transit. Hence, we conclude this is a false alarm signal.}
    \label{fig:toi225_bkg}
\end{figure}

\begin{figure*}
    \centering
    \includegraphics[width=\textwidth]{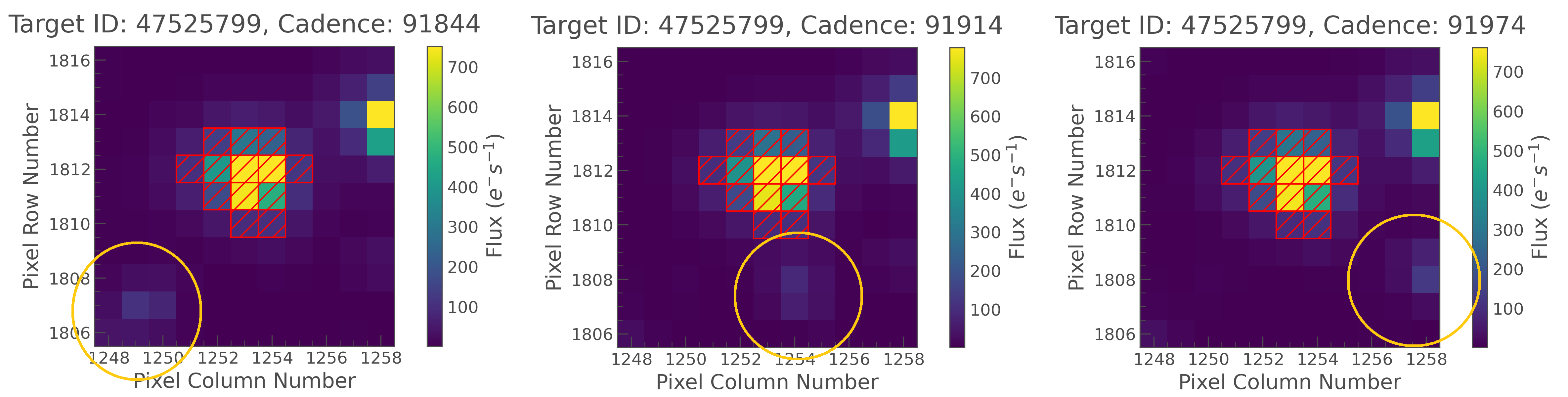}
    \caption{This plot shows the original CCD pixel TESS observations of TOI 225 from which the light curves of Figure \ref{fig:toi225_bkg} have been extracted. The red contour represents the aperture mask used by the SPOC pipeline. The left panel shows the FoV $\approx 2.3$ hours before the transit; the middle panel shows the FoV at the time of the transit; the right panel shows the FoV $\approx 2.3$ hours after the transit. The orange circle in each panel tracks the presence of a bright object passing in front of the camera. In the transit time-frame, the brightness of this object contaminates the pixels within the aperture mask used to extract the light curve thus mimicking a transit during the pre-processing of the signal.}
    \label{fig:toi225_bkg_2}
\end{figure*}

The mono-transit planet candidate TOI 225.01 detected by the SPOC turned out to be a false alarm due to the passage of a solar-system object (maybe an asteroid) in the field-of-view of TESS. TOI 225 is an early K star observed by TESS in sectors 2 and 29 in 2018 and 2020, respectively. The SPOC pipeline detected a mono-transit event at $\approx 1355.119$  BTJD with $S/N\approx 22.75$. When inspecting whether any background anomalies could have led to a false-alarm detection we found that is the case, as shown in Figs. \ref{fig:toi225_bkg} and \ref{fig:toi225_bkg_2}. 
TOI 225.01 is not an isolated occurrence within our dataset; strikingly, we have observed a similar event in the case of TOI 2456.01.
Thus we removed these mono-transit events as they originated from non-planetary scenarios.

\begin{figure*}
    \centering
    \includegraphics[width=\textwidth]{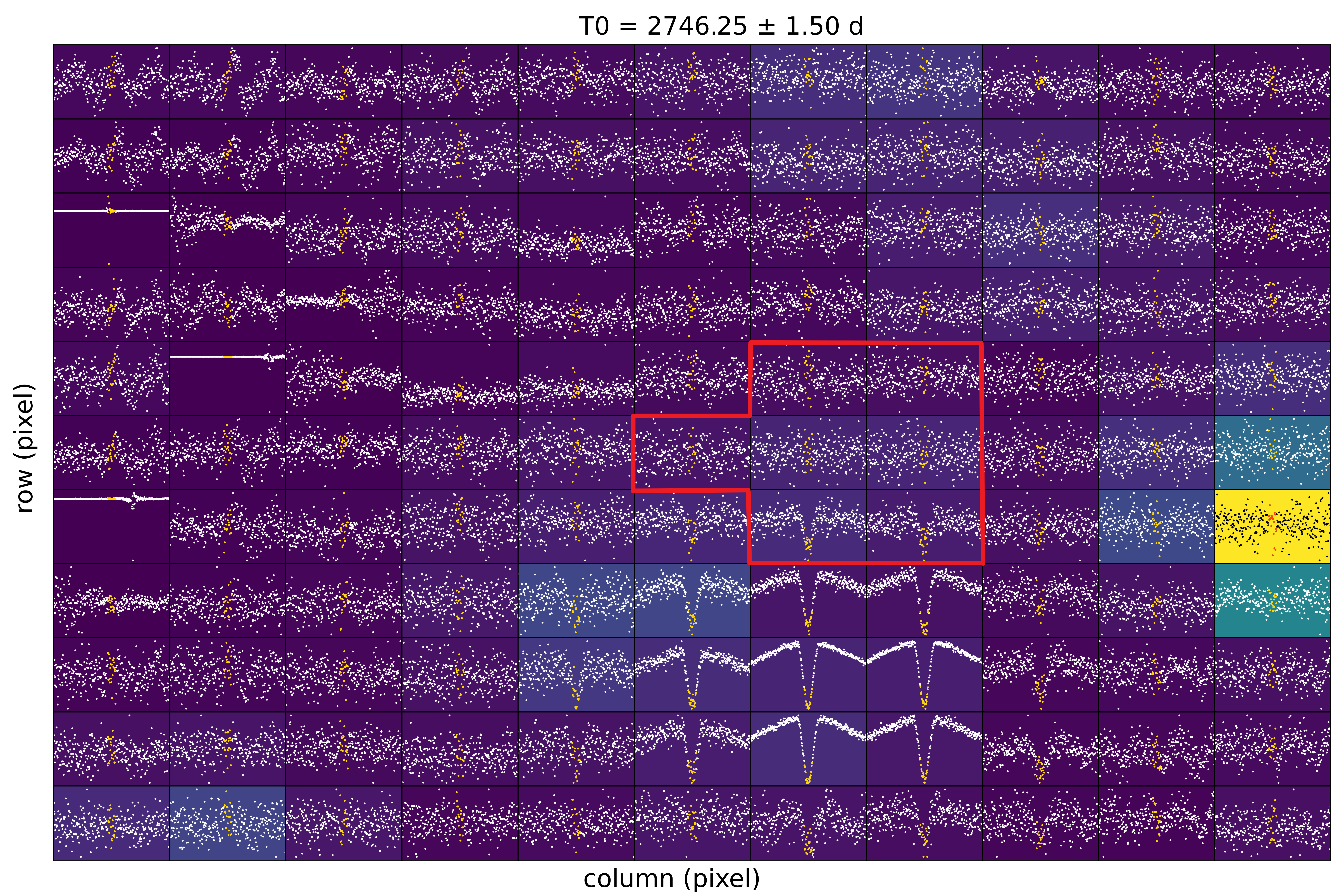}
    \caption{The Pixel Level light curve plot for planet candidate TOI 6092.01. It shows the light curve for each individual pixel of the corresponding target pixel file in a 3-day time-window centered at the time of transit $T0$.  Clear eclipses are seen in the bottom right pixels near the lower edge of the aperture mask (red contour). Hence, the source of the transit is not TOI 6092.01.}
    \label{fig:PLL_TOI6092}
\end{figure*}

In addition, we excluded the planet candidate TOI 6092.01 from our sample due to a distinct centroid offset observed during the transit event. As illustrated in Fig. \ref{fig:PLL_TOI6092}, the Pixel Level light curve exhibits a transit-like feature not on the pixel corresponding to our target, but rather on a neighboring pixel. Consequently, we ruled out TOI 6092.01 as a false positive due to the presence of a nearby transiting system within the aperture mask. Furthermore, the transit's V-shaped profile and significant depth suggest the possibility of it being a nearby eclipsing binary system. For the same reason, we also removed TOI 2007.01 from the list.

Despite V-shaped transits might be caused by grazing transiting exoplanets, they most likely are eclipsing binary systems. In addition, the best-fitting trapezoid model is inadequate for capturing the characteristics of a V-shaped transit. Due to their pronounced V-shaped transit profiles, we opted to exclude TOI 1581.01 and 2009.01.

\begin{figure}
    \centering
    \includegraphics[width=0.47\textwidth]{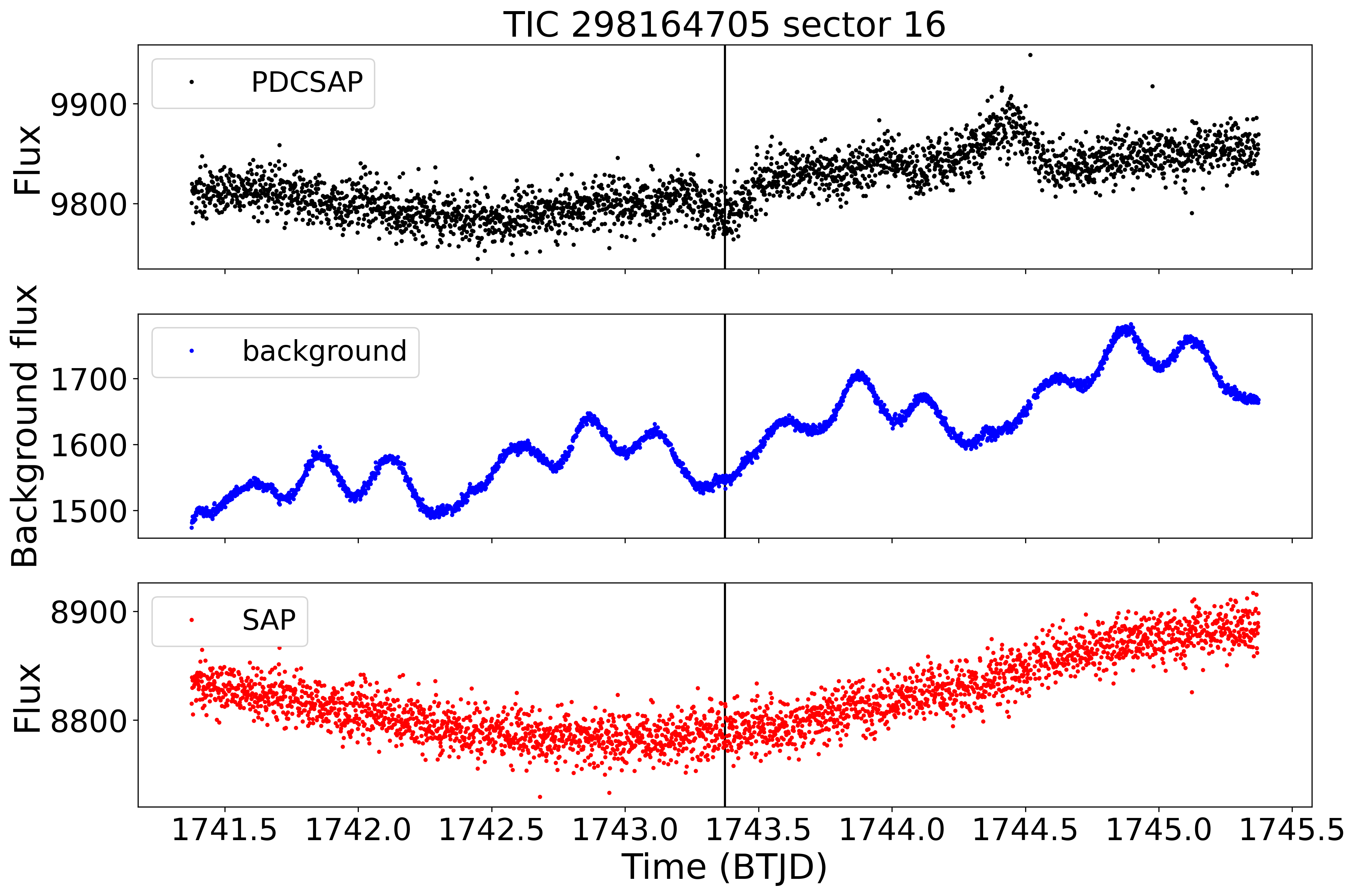}
    \caption{TOI 2435.01 is a mono-transit event detected by the SPOC pipeline in the TESS sector 16 around TOI 2435 (or TIC 298164705). In \textit{top panel} the PDCSAP light curve shows a transit at $\approx 1743.3$ BTJD, which completely vanishes in the SAP light curve in the \textit{bottom panel}. In the \textit{middle panel}, the background flux exhibits peculiar ringing modulations. We not only noticed the transit in the PDCSAP light curve occurred in correspondence with a dim in the background flux but also it has a fictitious spike at $\approx 1744.5$. We suspect that TOI 2435.01 could be a result of an aggressive pre-processing procedure that created an artificial transit-like feature in the PDCSAP light curve.}
    \label{fig:SAP_PDCSAP}
\end{figure}

Finally, we also removed from our sample TOI 1301.02, TOI 2087.01, TOI 2270.01, and TOI 2435.01 for a common reason. In each case, we observed a transit-like feature in the Pre-search Data Conditioning Simple Aperture Photometry (PDCSAP) light curve that disappeared upon verifying the Simple Aperture Photometry (SAP) light curve \citep{Kinemuchi2012}. Furthermore, while the SAP light curves do not exhibit any significant transit, the PDCSAP light curves display a noticeable correlation with background flux modulation as shown in Fig. \ref{fig:SAP_PDCSAP}. Consequently, we eliminated these suspicious cases from our sample to ensure we were working with the cleanest dataset possible.

\section{Results}
\label{sec:results}
Among the $48$ mono-transit TESS candidates, \npcs have successfully passed our rigorous vetting procedure, thus having a promising likelihood of confirmation by ground- or space-based missions.
We applied the technique discussed in the Sect. \ref{sec:methods} to our sample of \npcs mono-transit candidates in order to retrieve an estimate of their orbital period. While effective for high SNR transits, the simple trapezoid model is not intended for low SNR transit signals which are dominated by systematic noise sources (e.g., stellar activity, instrumental effects, flux contamination) and usually affect smaller planets orbiting larger stars. Gaussian Process (GP) modeling offers a viable alternative to the best-fitting trapezoid model due to its capability in capturing the effect of stellar variability during transit observations \citep{Rasmussen2006}. However, performing a comprehensive fitting of all the transit events using GP processes is computationally demanding and time-consuming. Hence, as a pragmatic approach, in this work we have chosen to exclude low SNR signals from our sample that cannot be accurately fitted by a simple trapezoid model. Table \ref{tab:results} presents the outcomes of our approach applied to the subset of $30$ mono-transit events from our sample that could be effectively fitted using a trapezoid model.

\begin{table}
\begin{tabular}{llll}
\hline
TOI     & $R_p (R_\oplus)$ & Period (days) & $i(^\circ)$ \\ \hline
289.01  & $16.42 \pm 0.73$  & $28 \pm 10$  &     $87.7 \pm 0.4$                             \\
429.01  & $9.03 \pm 0.22$  &   $415 \pm 94$ &  $89.83 \pm 0.01$                        \\
772.02  & $7.16 \pm 0.21$  & $55 \pm 28$   &    $89.40 \pm 0.07$                              \\
1386.01 & $6.83 \pm 0.20$  & $180 \pm 74$  &    $89.61 \pm 0.08$                              \\
1563.01 & $5.25 \pm 0.11$  & $242 \pm 57$  &    $89.78\pm 0.02$                              \\
1772.02 & $6.15 \pm 0.12$  & $120 \pm 44$   &   $89.55 \pm 0.07$                               \\
1812.01 & $9.18 \pm 0.18$  &  $208 \pm 48$  &    $89.76 \pm 0.01$                              \\
2085.01 & $3.72 \pm 0.15$  &  $551 \pm 454$ &    $89.9 \pm 0.1$                              \\
2134.02 & $8.02 \pm 0.17$  & $101 \pm 19$   &    $89.63 \pm 0.02$                              \\
2300.03 & $9.06 \pm 0.19$  &  $2736 \pm 424$  &    $89.96 \pm 0.02$                              \\
2423.01 & $14.03 \pm 0.24$  &$28 \pm 4$    &    $88.58 \pm 0.08$                              \\
2433.01 & $2.21\pm 0.27$  & $583 \pm 456$  &    $89.95\pm 0.05$                            \\
2436.01 & $2.66 \pm 0.15$  & $1302 \pm 1234$   &    $89.8 \pm 0.2$                              \\
2447.01 & $9.63 \pm 0.19$  & $84 \pm 19$   &    $89.8 \pm 0.2$                              \\
2490.01 & $11.7 \pm 0.4$  & $108 \pm 42$  &    $89.53\pm 0.03$                                \\
2529.01 & $11.55 \pm 0.36$  &$43 \pm 20$    &   $89.4 \pm 0.6$                               \\
2534.01 & $12.40 \pm 0.25$  & $39 \pm 15$   &    $89.5 \pm 0.5$                              \\
3724.01 & $14.14 \pm 0.52$  & $45 \pm 22$   &    $88.7 \pm 0.2$                              \\
4326.01 & $2.71 \pm 0.13$  & $142 \pm 134$   &   $89.2 \pm 0.7$                               \\
4465.01 & $11.83 \pm 0.19$  &$253 \pm 50$    &  $89.88 \pm 0.05$                                \\
4862.01 & $10.20 \pm 0.27$  & $113 \pm 31$   &   $89.54 \pm 0.05$                               \\
5523.01 & $11.11 \pm 0.23$  & $742 \pm 73$   &   $89.98 \pm 0.02$                               \\
5563.01 & $8.76 \pm 0.79$  & $78 \pm 29$   &    $89.2 \pm 0.2$                              \\
5564.01 & $11.77 \pm 0.42$  &$262 \pm 107$    &   $89.76 \pm 0.03$                               \\
5626.01 & $9.86 \pm 0.20$  & $9 \pm 2$   &      $87.4 \pm 0.3$                            \\
5719.01 & $2.59 \pm 0.21$  &  $1667 \pm 1586$  &   $89.8 \pm 0.2$                               \\
5745.01 & $1.07 \pm 0.08$  & $853 \pm 773$   &  $89.91 \pm 0.09$ \\
5893.01 & $10.44 \pm 0.36$  &$104 \pm 49$    &    $89.4 \pm 0.1$                              \\
5967.01 & $5.57 \pm 0.28$  & $368 \pm 176$   &    $89.71 \pm 0.08$                              \\
5968.01 & $3.23 \pm 0.17$  & $573 \pm 533$   &    $89.75 \pm 0.24$                              \\ \hline
\end{tabular}
\caption{The results of our methodology detailed in Sect. \ref{sec:methods} on 30 TOIs that we were able to fit with a simple trapezoid model.}
\label{tab:results}
\end{table}

\begin{figure*}
    \centering
    \includegraphics[width=\textwidth]{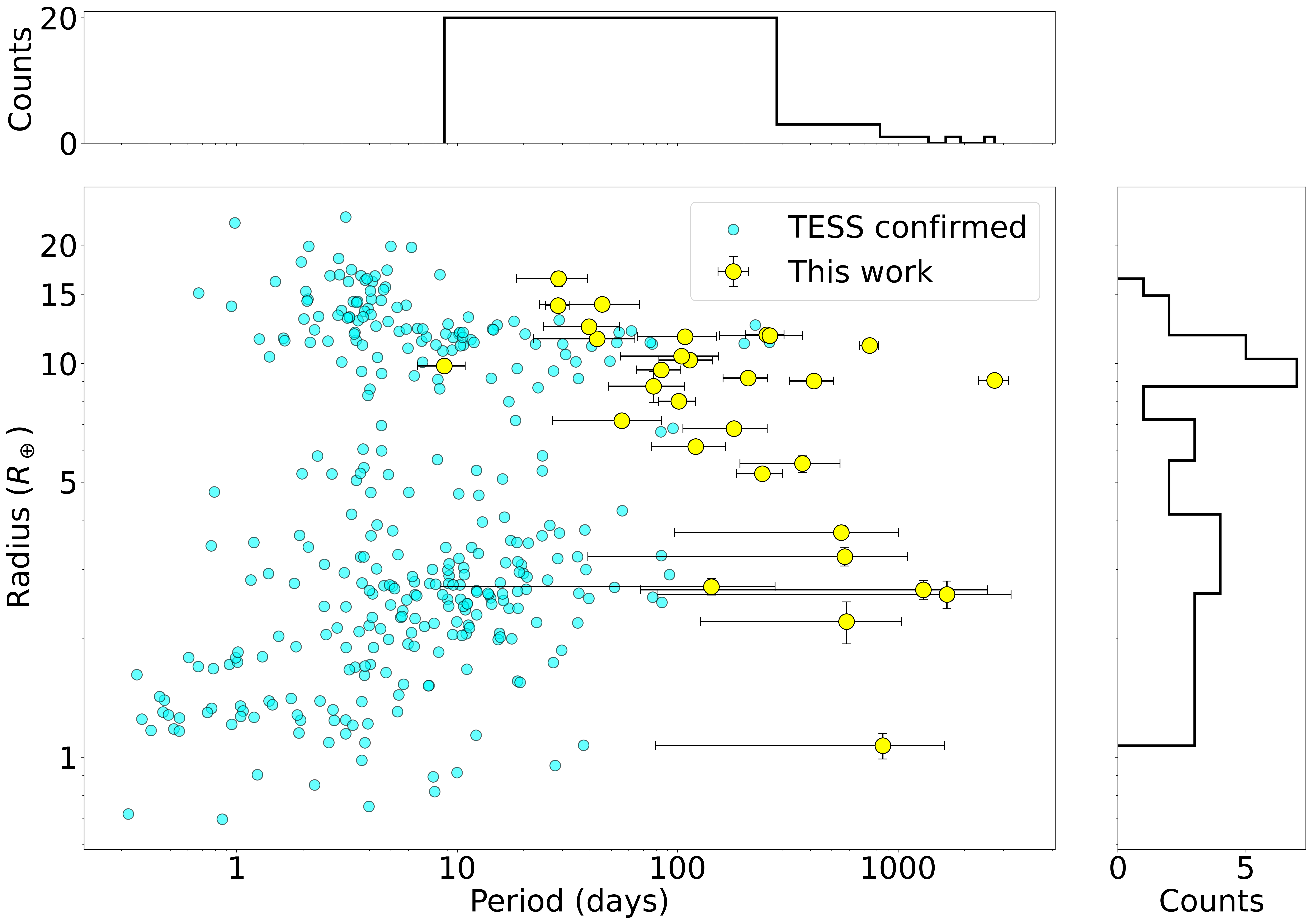}
    \caption{Distribution of the 30 mono-transit candidates within the $(P,R)$ diagram alongside their uncertainties over the orbital period and planetary radius. The cyan dots represent all the confirmed planets discovered by TESS.}
    \label{fig:PR_distribution}
\end{figure*}

We show the distribution of these 30 mono-transit planet candidates within the $(P, R)$ diagram in Fig. \ref{fig:PR_distribution}. 
What stands out in this figure is that our mono-transit planet candidates fill a region of the $(P, R)$ space-parameter that is mostly devoid of TESS confirmed exoplanets. The majority of candidates in our sample consist of warm and cold gas giants. 
As already discussed in Sect. \ref{sec:intro}, this was expected due to the TESS observing strategy. We also observe that the uncertainty over the orbital period is higher for the sub-Neptunes candidates in our sample, highlighting that our method loses its power for low SNR mono-transit events.

\begin{figure}
    \centering
    \includegraphics[width=0.47\textwidth]{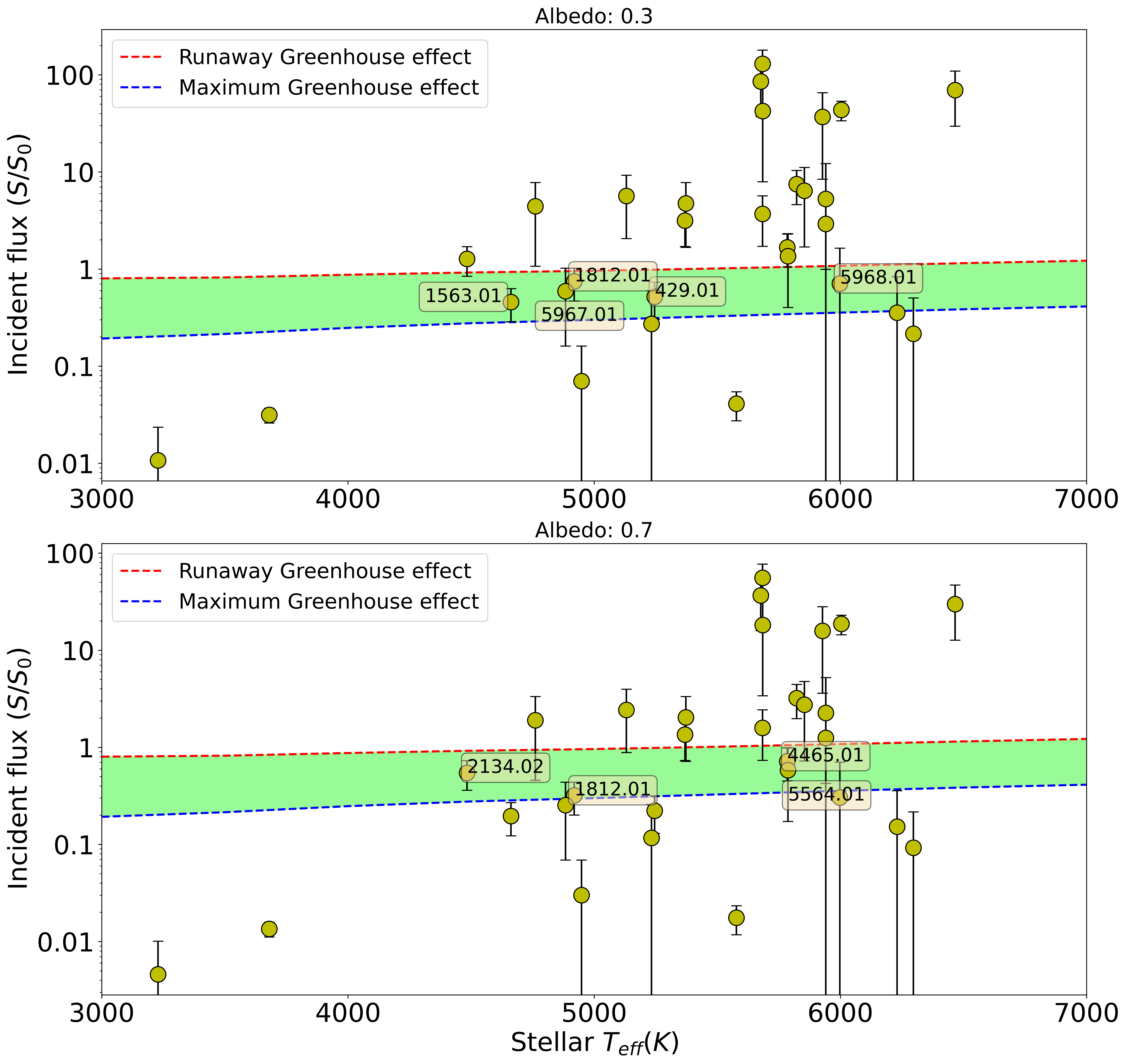}
    \caption{Distribution of the $30$ mono-transit TESS candidates (yellow dots) alongside the conservative habitable zone (green shaded area) of their host star for different values of the bond albedo $A$ ($A=0.3$ in the upper panel, $A=0.7$ in the bottom panel). The red and blue dashed lines represent respectively the runaway greenhouse effect and the maximum greenhouse effect edges \citep{Kopparapu2013}.}
    \label{fig:hz_candidates}
\end{figure}

For each of the 30 mono-transit candidates, we checked whether they reside within the habitable zone of their host stars. Specifically, we computed the circumstellar temperate zone boundaries based on the runaway greenhouse effect and the maximum greenhouse effect, as defined by \cite{Kopparapu2013} for different values of the planetary albedo. 
We found that TOI 2134.02, TOI 1812.01, TOI 5564.01 and TOI 4465.01 
potentially orbit within the circumstellar temperate zone when a high albedo value is assumed, as illustrated in Fig. \ref{fig:hz_candidates}. 
However, when considering lower albedo values similar to those found in the Solar System (e.g., 0.3), 
among the previous candidates 
only TOI 1812.01 is still found orbiting the circumstellar temperate zone. In addition, we also found that TOI 429.01, TOI 1563.01, TOI 5967.01 and TOI 5968.01 potentially lie within the circumstellar temperate zone. 
We highlight that out of the candidates potentially in the circumstellar temperate zone of their host star, TOI 429.01 is the only one that is additionally in the field of view of LOPS2.

\subsection{Interesting targets}
With a planetary radius of $R_p=16.42\pm 0.73 \, R_\oplus$, TOI 289.01 is the largest candidate in our sample. 
It revolves around its star in $28\pm 10$ days.
Even though there is a higher probability for very large transiting objects to be brown dwarfs rather than planets (i.e., the FP probability increases towards a bigger radius, see for example \citealt{Magliano2023Patrol}), TOI 289.01 did not show any clue of an eclipsing binary scenario during our photometric analysis. However, since TOI 289 is contained in the Gaia Non-Single stars catalogue, \cite{Tarrants2023} recently confirmed it as being a spectroscopic binary with a $P=224.706 \pm 0.906$ days. Using the orbital parameters obtained by \cite{Holl2023} from Gaia data, \cite{Tarrants2023} found that the single transit detected by TESS closely matches with the secondary eclipse of the system. Thus, the companion of TOI 289 is a $1.4R_J$ object on an eccentric orbit $e=0.35848\pm 0.08394$.

TOI 2300.03 is found to be the furthest candidate orbiting its star in $2736\pm 424$ days at the equilibrium temperature of $\approx 125 \, {\rm K}$, assuming a null planetary albedo. What makes this candidate particularly interesting is that it could belong to a multi-planetary system composed of a candidate super-Earth (TOI 2300.01) and a candidate Neptune (TOI 2300.02), both on close-in orbits, and a much further away Jupiter-like companion (TOI 2300.03). This system is a valuable planetary system to test models of planetary evolution. 
Finally, we remark that if TOI 2300.03 is on a high-eccentric orbit, its orbital period could be drastically reduced as we have seen for TOI 2180 b in Sect. \ref{sec:test}.

A similar scenario holds for TOI 5523.01, a Jupiter-like candidate that completes a full orbit around its star (K2-43) in  $742\pm 73$ days. Notably, K2-43 has been monitored during the first K2 campaign in 2014 \citep{Crossfield2016,Dressing2017,Hedges2019,Kruse2019}. We inspected the K2 light curve to check whether a previous transit event was registered in the past but the time spanning by the K2 observations did not overlap with the predicted transit of TOI 5523.01. 
It is important to mention that K2-43 is already recognized as the host of two confirmed exoplanets, K2-43 b and K2-43 c, on short-period orbits. Therefore, if TOI 5523.01 is confirmed as a \textit{bonafide} planet through subsequent observations as K2-43 d, it would become the third planet in this planetary system.  

The sub-Jovians TOI 772.02 and TOI 1772.02 2 also could form two-planet systems with either another sub-Jovian companion candidate (TOI 772.01) or a mini-Neptune candidate (TOI 1772.01).  

\cite{Rescigno2023} have recently confirmed that the candidate TOI 2134.02 is a sub-Saturn planet orbiting its K-dwarf star in $95.50^{+0.36}_{-0.25}$ days on a highly eccentric orbit ($e=0.64^{+0.05}_{-0.06}$). We here stress that our $101\pm 19$ days prediction, which comes from the $P_2$-branch, deviates from the actual orbital planet by $6\%$ due to its eccentric orbit. The obtained overestimate is consistent with our findings discussed in Sect. \ref{sec:beyond_circular}.

TOI 1812.01 is a $9.18\, R_\oplus$ candidate on a $(208\pm 48)$ days orbit that could form a three-planet system with its candidate companions, the mini-Neptune TOI 1812.02 and the Neptune-sized TOI 1812.03 which orbit their host star in $\approx 11$ and $\approx 48$ days, respectively.

The giant TOI 5564.01 ($R_p = 11.77 \pm 0.42$)  orbiting its star in $262\pm 107$ days could be the outer companion of the planet candidate TOI 5564.02, a sub-Jupiter planet on a $45$ days orbit.

Similarly, TOI 5968.01 ($R_p=3.23\pm 0.17$) completing an orbital in $573$ days ($\approx 93\%$ uncertainty) could form a two-planet system with the mini-Neptune planet candidate TOI 5968.02 which completes its orbit in $\approx 9$ days.

TOI 5626.01 has been detected by the SPOC pipeline at $\approx 2660.04$ BTJD in Sector 49 which spans a time-window of $[2637.47,2664.32]$ BTJD with a gap in $[2651.42,2655.83]$ BTJD. This is a puzzling candidate because we found it orbits its variable host star in $9\pm 2$ days, thus it should have been transited at least once in the time-frame $[2637.47,2651.42]$ BTJD. We actually found a possible additional transiting feature at $2643.96$ BTJD which unfortunately occurred just before a TESS momentum dump making the measurement unreliable. However, it is worth noting that TOI 5626 was also observed by TESS in Sector 23, during which no evident transiting-like feature was observed. These circumstances raise doubts regarding the true nature of TOI 5626.01. Additionally, although TOI 5626.01 successfully passed our vetting procedure, it is important to consider the possibility that it may possess a non-negligible eccentricity. Such eccentricity can significantly impact the outcomes and interpretations derived from our approach.

A similar case is TOI 2423.01, detected by the SPOC pipeline in Sector 29 at $\approx 2097.06$ BTJD. Our analysis determined an orbital period of $28\pm 4$ days. However, despite further observations of the star in Sectors 2, 4, 30, and 31, no additional analogous transits were observed. Hence, we are quite suspicious about the true nature of TOI 2423.01 but it could also be revolving around its star on an eccentric orbit.

In a recent work, \cite{Hawthorn2023} found one additional transit for both TOI 2490.01 and TOI 2529.01, by performing a systematic duotransit search. Following the same methodology discussed in \cite{Osborn2022b} they found the most probable orbital period for TOI 2490.01 to be $P=38.01^{+14.0}_{-15.0}$ days, while $P=65.0^{+24.0}_{-5.0}$ days for TOI 2529.01.

\subsection{Follow-up efforts}
We have also investigated the feasibility of follow-up observing campaigns on these targets using the CHaracterising ExOPlanets Satellite (CHEOPS, \citealt{Broeg2013,Fortier2014}) telescope. CHEOPS is an ESA space-based mission launched on December 18, 2019, whose main goal is to perform ultrahigh precision photometry on bright stars already known to host planets. 

\begin{figure}
    \centering
    \includegraphics[width=0.47\textwidth]{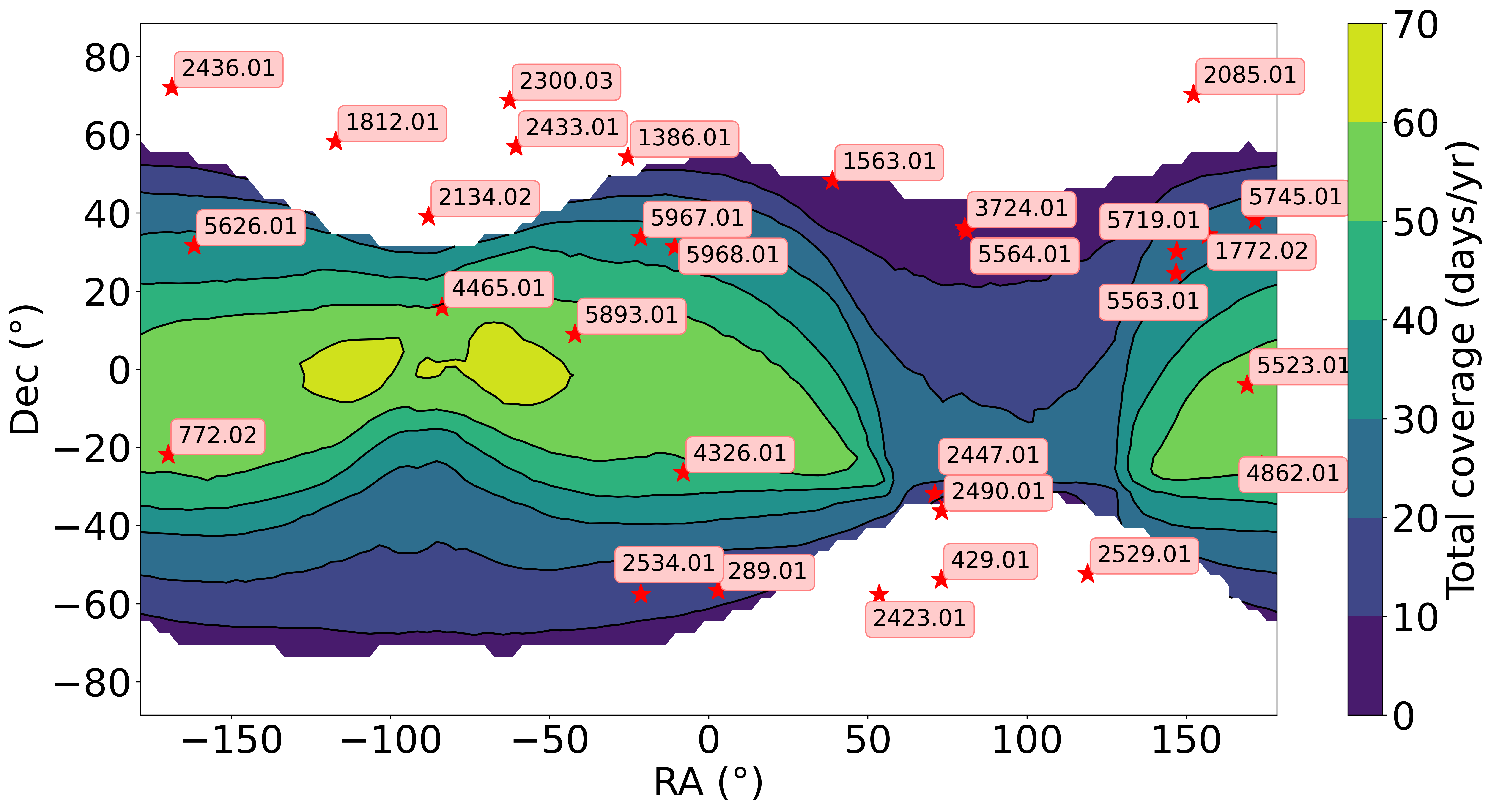}
    \caption{The $30$ TESS mono-transit planet candidates in our sample (represented by red star symbols) on the CHEOPS sky coverage map using an observing efficiency of $\approx 60\%$. Each star that falls in the contours would be monitored by CHEOPS for a minimum of $59$ minutes per orbit. The color bar represents the total observing time in days using CHEOPS over a period of 1 year.}
    \label{fig:CHEOPS_vis}
\end{figure}

Thus we checked whether the $30$ mono-transit planet candidates in our sample would be amenable targets of a CHEOPS' observer programme following the steps of \cite{Cooke2020}. Due to its specific orbit around the Earth, observations from CHEOPS experience interruptions along its trajectory leading to different sky coverage depending on the requested time per CHEOPS orbit (\textit{observing efficiency}). 
The optimal requested time per orbit is given by the best trade-off between maximizing both the sky and temporal coverage.
In this work, we opted for a minimum observing time of $59$ minutes per orbit (equivalent to approximately $60\%$ of the observing efficiency) as it matches quite well with the typical ingress/egress duration of our candidates.  
We plot the distribution of the $30$ mono-transit planet candidates over the CHEOPS sky coverage using an observation efficiency of $60\%$ in Fig. \ref{fig:CHEOPS_vis}. As we can see from this chart, almost $36\%$ of our final sample falls within the uncovered sky region. Moreover, the most promising stars that CHEOPS would monitor for $50\div 60$ days in $1$ year are TOI $772$, TOI $4465$, TOI $5893$, TOI $523$ and TOI $4862$. As found by \cite{Cooke2020}, we also highlight that none of our TESS mono-transit candidates lie in the region of the sky CHEOPS monitors for the longest. 

Despite the feasibility to schedule photometric follow-up campaigns, we acknowledge that RV follow-up observations could significantly accelerate the determination of orbital periods for mono-transit candidates. In comparison to photometric observations, RV measurements may give the advantage of acquiring a data point for a single candidate in a single observing night. Thus, the addition of RV follow-up programmes would significantly improve the efficiency of our mono-transit candidate validation procedure.

\section{Conclusions}
\label{sec:conclusions}

In this study, we have investigated the mono-transit events observed by the TESS mission around stars that will be further monitored by the PLATO mission. We developed an analytic approach that enables us to estimate the orbital period and inclination of a mono-transit signal by exploiting only the shape of the single transit. 
The effectiveness of the method proposed in this work depends on both the photometric accuracy of the measurements and the uncertainty associated with the stellar parameters under the assumption of a circular orbit. 
We applied the method on a sample of $38$ uniformly vetted TESS mono-transit planet candidates orbiting stars that could be potentially PLATO targets, out of the initial 48 detected signals. We constrained the orbital period and inclination for $30$ objects in our sample. These planet candidates encompass a region of the space of parameters still unexplored by the TESS mission. Among these targets, at least four will be observed on the first LOP field, LOPS2, starting from the end of 2026.
Since our vetting procedure is based only on the single observed transit, these candidates are still away from a robust validation, thus follow-up observations are needed to validate and properly characterize them. We also checked whether the candidates in our sample could be amenable targets for a CHEOPS observing campaign. 

Through the cross-matching of TESS data with the PLATO Input Catalog, we have demonstrated a potential synergy between the two missions. As TESS continues to operate and collect data, its discoveries set the stage for further investigations and pave the way for future missions, such as PLATO. The availability of PLATO observations will play a crucial role in validating the planetary nature of mono-transit candidates and discerning them from false positives, such as blended stars or background eclipsing binaries. 

As the PLATO mission continues to observe the stars monitored by TESS, we anticipate significant advancements in the confirmation, characterization, and statistical analysis of mono-transit events.

\section*{Acknowledgements}
We would like to thank the anonymous referee for their careful reading and their insightful comments which contributed to improve the quality of this manuscript.

This research has made use of the NASA Exoplanet Archive, which is operated by the California Institute of Technology, under contract with the National Aeronautics and Space Administration under the Exoplanet Exploration Program.

We acknowledge the use of the Exoplanet Follow-up Observation Program website, which is operated by the California Institute of Technology, under contract with the National Aeronautics and Space Administration under the Exoplanet Exploration Program. The TFOP is led by the Smithsonian Astrophysical Observatory (SAO), in coordination with MIT, as part of the TESS Science Office.

This work has made use of data from the European Space Agency (ESA) mission {\it Gaia} (\url{https://www.cosmos.esa.int/gaia}), processed by the {\it Gaia}
Data Processing and Analysis Consortium (DPAC, \url{https://www.cosmos.esa.int/web/gaia/dpac/consortium}). Funding for the DPAC has been provided by national institutions, in particular the institutions participating in the {\it Gaia} Multilateral Agreement. JSJ greatfully acknowledges support by FONDECYT grant 1201371 and from the ANID BASAL projects ACE210002 and FB210003. JIV acknowledges support of CONICYT-PFCHA/Doctorado Nacional-21191829.

\textit{Software}: \texttt{topcat} \citep{Taylor2005}, lightkurve \citep{Lightkurve2018}, \texttt{ARIADNE} \citep{Vines2022}.

\section*{Data Availability}
The data underlying this article are taken from the literature.



\bibliographystyle{mnras}
\bibliography{biblio} 




\appendix
\section{Formal solutions to polynomial equation} 
\label{appendix: formal_solutions}

In this Appendix we show the key steps to find the analytic solutions to equation \eqref{eq:eq_param} that here we write again
\begin{equation}
    \alpha P^2-\beta P^{2/3}+\gamma=0.
    \label{eq1}
\end{equation}
Initially, we introduce the notation $P^{2/3}\equiv y$ to conveniently reframe equation \eqref{eq1} in the subsequent manner,
\begin{equation}
    \alpha y^3-\beta y+\gamma=0\quad\Leftrightarrow\quad y^3+py+q=0,
    \label{eq2}
\end{equation}
where we set $p\equiv -\beta/\alpha$ and $q\equiv\gamma/\alpha.$
The equation \eqref{eq2} is a cubic equation whose solutions can be found using the so called \textit{Cardano}'s formula. We first define the discriminant $\Delta$ as
\begin{equation}
    \Delta \equiv \dfrac{q^2}{4}+\dfrac{p^3}{27}=\dfrac{1}{\alpha^2}\left(\dfrac{\gamma^2}{4}-\dfrac{\beta^3}{27\alpha}\right).
    \label{def_delta}
\end{equation}
Depending on the value of the discriminant $\Delta$, we have the following cases.
    \paragraph*{Case 1 - $\Delta > 0$.\newline} 
    
    In this case we have three different solutions to equation \eqref{eq2}:
    \begin{align}
    y_1 &= u+v, \\
    y_2 &=u\left(-\dfrac{1}{2}+i\dfrac{\sqrt{3}}{2}\right)+v\left(-\dfrac{1}{2}-i\dfrac{\sqrt{3}}{2}\right),  \\
    y_3 &= u\left(-\dfrac{1}{2}-i\dfrac{\sqrt{3}}{2}\right)+v\left(-\dfrac{1}{2}+i\dfrac{\sqrt{3}}{2}\right), 
    \end{align}
    where 
    \begin{equation}
        u=\left(-\dfrac{q}{2}+\sqrt{\Delta}\right)^{1/3}\quad\text{and}\quad v=\left(-\dfrac{q}{2}-\sqrt{\Delta}\right)^{1/3} \, .
    \end{equation}

We note that among the solutions $y_1$, $y_2$, and $y_3$, only $y_1$ is real and has a physical meaning. Nevertheless, as $p\equiv -\beta/\alpha<0$, it becomes evident that both $u$ and $v$ assume negative values. Consequently, $y_1<0$, but since the physical variable of interest is the orbital period denoted by $P\equiv y^{3/2}$, we can conclude that equation \eqref{eq1} does not yield a viable physical solution in this particular scenario.

\paragraph*{Case 2 - $\Delta < 0$.\newline}
  
In this scenario, it is necessary to compute the true anomaly $\theta$ of the complex number $\left(-q/2+i\sqrt{-\Delta}\right)$, which can be determined using the following expression:
    \begin{equation}
        \theta=\pi-\arctan\left(2\sqrt{-\Delta}/q\right).
        \label{anomaly}
    \end{equation}
Then there are three solutions to equation \eqref{eq2},
    \begin{align}
    y_1 &= 2\sqrt{\dfrac{-p}{3}}\cos\left(\dfrac{\theta}{3}\right), \\
    y_2 &=2\sqrt{\dfrac{-p}{3}}\cos\left(\dfrac{\theta+2\pi}{3}\right),  \\
    y_3 &= 2\sqrt{\dfrac{-p}{3}}\cos\left(\dfrac{\theta+4\pi}{3}\right). 
    \end{align}
    As $\arctan(2\sqrt{-\Delta}/q)\in \left[0,\pi/2\right]$, it follows that the true anomaly $\theta\in\left[\pi/2,\pi\right]$. Consequently, we can readily observe that only solutions $y_1$ and $y_3$ yield positive values, while $y_2<0$. By substituting the expression of $\theta$ into $y_1$ and $y_3$, we can derive the following results:
    \begin{align}
    y_1 &= 2\sqrt{\dfrac{-p}{3}}\cos\left(\dfrac{\pi-\arctan\left(2\sqrt{-\Delta}/q\right)}{3}\right), \\
    y_3 &=2\sqrt{\dfrac{-p}{3}}\cos\left(\dfrac{5\pi-\arctan\left(2\sqrt{-\Delta}/q\right)}{3}\right).
    \end{align}
    Then one can retrieve the formula for the orbital periods $P_1=y_3^{3/2}$ and $P_2=y_1^{3/2}$ shown in the Sect.\ref{sec:polynomial_eq}, thus  
    \begin{align}    
    P_1&=2^{3/2}\left(\dfrac{\beta}{3\alpha}\right)^{3/4}\cos^{3/2}\Bigg[\dfrac{5}{3}\pi-\dfrac{1}{3}\arctan\left(\sqrt{\dfrac{4\beta^3}{27\alpha\gamma^2}-1}\right)\Bigg],\\
    P_2&=2^{3/2}\left(\dfrac{\beta}{3\alpha}\right)^{3/4}\cos^{3/2}\Bigg[\dfrac{\pi}{3}-\dfrac{1}{3}\arctan\left(\sqrt{\dfrac{4\beta^3}{27\alpha\gamma^2}-1}\right)\Bigg] \, .
    \end{align}
    The uncertainties $\Delta P_1$ and $\Delta P_2$ associated respectively to $P_1$ and $P_2$ are given by
    \begin{align}
        \Delta P_1&=P_1(\sigma_{\alpha,1}+\sigma_{\beta,1}+\sigma_{\gamma,1}),\\
        \Delta P_2&=P_2(\sigma_{\alpha,2}+\sigma_{\beta,2}+\sigma_{\gamma,2}),
    \end{align}
    where we posed
    \begin{align}
        \sigma_{\alpha,1}&\equiv\dfrac{\Delta \alpha}{4\alpha}\Biggl\{3+\dfrac{\sqrt{\Delta}}{\Delta+1}\Bigg|\tan\Bigg[\dfrac{5}{3}\pi-\dfrac{1}{3}\arctan\left(\sqrt{\dfrac{4\beta^3}{27\alpha\gamma^2}-1}\right)\Bigg]\Bigg|\Biggr\},\\
        \sigma_{\alpha,2}&\equiv\dfrac{\Delta \alpha}{4\alpha}\Biggl\{3+\dfrac{\sqrt{\Delta}}{\Delta+1}\Bigg|\tan\Bigg[\dfrac{\pi}{3}-\dfrac{1}{3}\arctan\left(\sqrt{\dfrac{4\beta^3}{27\alpha\gamma^2}-1}\right)\Bigg]\Bigg|\Biggr\},\\
        \sigma_{\beta,1}&\equiv\dfrac{3\Delta \beta}{4\beta}\Biggl\{1+\dfrac{\sqrt{\Delta}}{\Delta+1}\Bigg|\tan\Bigg[\dfrac{5}{3}\pi-\dfrac{1}{3}\arctan\left(\sqrt{\dfrac{4\beta^3}{27\alpha\gamma^2}-1}\right)\Bigg]\Bigg|\Biggr\},\\
        \sigma_{\beta,2}&\equiv\dfrac{3\Delta \beta}{4\beta}\Biggl\{1+\dfrac{\sqrt{\Delta}}{\Delta+1}\Bigg|\tan\Bigg[\dfrac{5}{3}\pi-\dfrac{1}{3}\arctan\left(\sqrt{\dfrac{4\beta^3}{27\alpha\gamma^2}-1}\right)\Bigg]\Bigg|\Biggr\},\\
        \sigma_{\gamma,1}&\equiv\dfrac{\Delta\gamma}{2\gamma}\dfrac{\sqrt{\Delta}}{\Delta+1}\Bigg|\tan\Bigg[\dfrac{5}{3}\pi-\dfrac{1}{3}\arctan\left(\sqrt{\dfrac{4\beta^3}{27\alpha\gamma^2}-1}\right)\Bigg]\Bigg|,\\
        \sigma_{\gamma,2}&\equiv\dfrac{\Delta\gamma}{2\gamma}\dfrac{\sqrt{\Delta}}{\Delta+1}\Bigg|\tan\Bigg[\dfrac{\pi}{3}-\dfrac{1}{3}\arctan\left(\sqrt{\dfrac{4\beta^3}{27\alpha\gamma^2}-1}\right)\Bigg]\Bigg|,\\
        \label{eq:errors}
    \end{align}
    with $\Delta\alpha,\Delta\beta$ and $\Delta\gamma$ the uncertainties on the parameters $\alpha,\beta$ and $\gamma$ respectively.
    \paragraph*{Case 3 - $\Delta = 0$.\newline}
    We address this particular scenario to ensure comprehensive coverage; however, it is important to note that in practical situations, we would not encounter an exact value of $\Delta =0$. Nonetheless, even in this hypothetical case, equation \eqref{eq2} still yields three solutions.
    \begin{align}
        y_1 &=-2\left(\dfrac{q}{2}\right)^{1/3},\\
        y_2 &=y_3=\left(\dfrac{q}{2}\right)^{1/3}.
    \end{align}
    We rule out the negative solution $y_1$ and so equation \eqref{eq1} admits a single solution
    
    \begin{equation}
    P=y_2^{3/2}=\sqrt{\dfrac{q}{2}}=\sqrt{\dfrac{\gamma}{2\alpha}} \, .
    \end{equation}

\section{Equations and approximations}
\label{app:summary}
In this Appendix we make a summary of the all equations and approximations we used in this work alongside, where available, their analytic solutions.

\begin{table*}
    \centering
    
    \begin{tabular}{|c|c|c|}
        \hline
        \textbf{Approximation} & \textbf{Simplified Equation} & \textbf{Analytic solutions} \\
        \hline
        $\begin{aligned} 
        &\text{Circular orbit}\\
        &\text{No flux contamination}\\
        &\text{No limb-darkening}
        \end{aligned}$ & $t_T=\dfrac{P}{\pi}\arcsin{\left(\dfrac{R_*}{a}\left[\dfrac{[1+\sqrt{\delta}]^2-[a\cos i/R_*]^2}{1-\cos ^2 i}\right]^{1/2}\right)}$ & -  \\
        \hline
        $M_p\ll M_*$ & $t_T=\dfrac{P}{\pi}\arcsin{\left[\left(\dfrac{\Lambda^{-2/3}_{*}P^{-4/3}(1+\sqrt{\delta})^2-\cos^2(i)}{1-\cos^2(i)}\right)^{1/2}\right]}$ & - \\
        \hline
        $\begin{aligned} &\pi t_T/P\ll 1\\ &\cos(i)\ll 1 \end{aligned}$ & $\cos^2(i) \, P^2-\Lambda^{-2/3}_{*}(1+\sqrt{\delta})^2 \, P^{2/3}+\pi^2 \, t_{T}^2=0$ & $ \begin{aligned} P_1=&2^{3/2}\left(\dfrac{\beta}{3\alpha}\right)^{3/4}\cos^{3/2}\Bigg[\dfrac{5}{3}\pi-\dfrac{1}{3}\arctan\left(\sqrt{\dfrac{4\beta^3}{27\alpha\gamma^2}-1}\right)\Bigg]\\ P_2=&2^{3/2}\left(\dfrac{\beta}{3\alpha}\right)^{3/4}\cos^{3/2}\Bigg[\dfrac{\pi}{3}-\dfrac{1}{3}\arctan\left(\sqrt{\dfrac{4\beta^3}{27\alpha\gamma^2}-1}\right)\Bigg]\end{aligned}$ \\
        \hline
        $i=90^\circ$ & $\Lambda^{-2/3}_{*}(1+\sqrt{\delta})^2 \, P^{2/3}-\pi^2 \, t_{T}^2=0$ & $P_{\text{EO}}=\left(\dfrac{\pi \, t_T}{1+\sqrt{\delta}}\right)^3\Lambda_*$\\
        \hline
        \textbf{Approximation} & \textbf{Simplified Equation} & \textbf{Analytic solutions} \\
        \hline
        $ \begin{aligned} &\text{Circular orbit}\\
                          &\text{No flux contamination}\\
                          &\text{No limb-darkening}\\
                         & M_p\ll M_*\\
        \end{aligned}$ & $\dfrac{M_*}{R_*^3}=\left(\dfrac{4\pi^2}{P^2G}\right)\Biggl\{\dfrac{(1+\sqrt{\delta})^2-b^2(1-\sin^2(t_T\pi/P)}{\sin^2(t_T\pi/P)}\Biggr\}^{3/2}$ & - \\
        \hline
        $\pi t_T/P\ll 1$ & $\dfrac{M_*}{R_*^3}=\dfrac{32\delta^{3/4}P}{G\pi(t_T^2-t_F^2)^{3/2}}$ & $P_{t_F}=\dfrac{G\pi^2}{24}\rho_*\dfrac{(t_T^2-t_F^2)^{3/2}}{\delta^{3/4}}$\\
        \hline
    \end{tabular}
    \caption{Summary of equations used in this work and their analytic solutions.}
    \label{tab:summary_table}
\end{table*}

\bsp	
\label{lastpage}
\end{document}